\documentclass[useAMS,usenatbib]{mn2e}
\usepackage{graphicx,epsfig}

\def\eg{{e.g.,~}}
\def\ie{{i.e.,~}}

\title[EBL Inferred from AEGIS Galaxy SED-type Fractions]{Extragalactic Background Light Inferred from AEGIS Galaxy SED-type Fractions}

\author[A. Dom\'inguez et al.] {A. Dom\'inguez$^{1,2,3}$\thanks{E-mail: alberto@iaa.es}, J.~R. Primack$^{4}$, D.~J. Rosario$^{5}$, F. Prada$^{2,6}$, R.~C. Gilmore$^{4,7}$,\newauthor S.~M. Faber$^{5}$, D.~C. Koo$^{5}$, R.~S. Somerville$^{8}$, M.~A. P\'erez-Torres$^{2}$,\newauthor P. P\'erez-Gonz\'alez$^{9,10}$, J.-S. Huang$^{11}$, M. Davis$^{12}$, P. Guhathakurta$^{5}$,\newauthor P. Barmby$^{13}$, C.~J. Conselice$^{14}$, J.~A. Newman$^{15}$, M.~C. Cooper$^{16}$ and M. Lozano$^{1}$\\
	$^1$Departamento de F\'isica At\'omica, Molecular y Nuclear, Universidad de Sevilla, Apdo. Correos 1065, E-41080 Sevilla, Spain\\
	$^2$Instituto de Astrof\'isica de Andaluc\'ia, CSIC, Apdo. Correos 3004, E-18080 Granada, Spain\\
	$^3$Visiting student at the Santa Cruz Institute for Particle Physics (SCIPP),  University of California, Santa Cruz, CA 95064, USA\\
	$^4$Department of Physics, University of California, Santa Cruz, CA 95064, USA\\
	$^5$UCO/Lick Observatory, Dept. of Astronomy \& Astrophysics, University of California, Santa Cruz, CA 95064, USA\\	
	$^6$Visiting researcher at the Santa Cruz Institute for Particle Physics (SCIPP),  University of California, Santa Cruz, CA 95064, USA\\
	$^7$Scuola Internazionale Superiore di Studi Avanzati (SISSA), Via Bonomea 265, 34136, Trieste, Italy\\
	$^8$Space Telescope Science Institute, 3700 San Martin Drive, Baltimore, MD 21218, USA\\
	$^9$Departamento de Astrof\'isica, Facultad de CC. F\'isicas, Universidad Complutense de Madrid, E-28040 Madrid, Spain\\
	$^{10}$Associate Astronomer at Steward Observatory, The University of Arizona\\
	$^{11}$Harvard-Smithsonian Center for Astrophysics, 60 Garden Street, Cambridge, MA 02138, USA\\	
	$^{12}$Department of Astronomy, University of California, Berkeley, CA 94720, USA\\
	$^{13}$Department of Physics \& Astronomy, University of Western Ontario, London, ON N6A 3K7, Canada\\
	$^{14}$University of Nottingham, School of Physics \& Astronomy, Nottingham NG7 2RD, UK\\
	$^{15}$Department of Physics and Astronomy, University of Pittsburgh, 3941 O'Hara Street, Pittsburgh, PA 15260\\
	$^{16}$Steward Observatory, University of Arizona, 933 N. Cherry Avenue, Tucson, AZ; Spitzer fellow}
	
\begin{document}
\label{firstpage}

\date{}

\pagerange{\pageref{firstpage}--\pageref{lastpage}} \pubyear{2010}

\maketitle

\begin{abstract}
The extragalactic background light (EBL) is of fundamental importance both for understanding the entire process of galaxy evolution and for $\gamma$-ray astronomy, but the overall spectrum of the EBL between 0.1~-~1000~$\mu$m has never been determined directly from galaxy spectral energy distribution (SED) observations over a wide redshift range. The evolving, overall spectrum of the EBL is derived here utilizing a novel method based on observations only. This is achieved from the observed evolution of the rest-frame $K$-band galaxy luminosity function up to redshift 4 (\citealt{cirasuolo10}), combined with a determination of galaxy SED-type fractions. These are based on fitting SWIRE templates to a multiwavelength sample of about 6000 galaxies in the redshift range from 0.2 to 1 from the All-wavelength Extended Groth Strip International Survey (AEGIS). The changing fractions of quiescent galaxies, star-forming galaxies, starburst galaxies and active galactic nucleus (AGN) galaxies in that redshift range are estimated, and two alternative extrapolations of SED-types to higher redshifts are considered. This allows calculation of the evolution of the luminosity densities from the UV to the IR, the evolving star formation rate density of the universe, the evolving contribution to the bolometric EBL from the different galaxy populations including AGN galaxies and the buildup of the EBL. Our EBL calculations are compared with those from a semi-analytic model, from another observationally-based model and observational data. The EBL uncertainties in our modeling based directly on the data are quantified, and their consequences for attenuation of very high energy $\gamma$-rays due to pair production on the EBL are discussed. It is concluded that the EBL is well constrained from the UV to the mid-IR, but independent efforts from infrared and $\gamma$-ray astronomy are needed in order to reduce the uncertainties in the far-IR.\\
\end{abstract}


\section{Introduction}
\label{intro}

The formation and evolution of galaxies in the universe is accompanied unavoidably by the emission of radiation. All this radiated energy is still streaming through the universe, although much is now at longer wavelengths due to redshifting and absorption/re-emission by dust. The photons mostly lie in the range $\sim$~0.1-1000~$\mu$m, \ie ultraviolet (UV), optical, and infrared (IR), and produce the second most energetic diffuse background after the Cosmic Microwave Background, thus being essential for understanding the full energy balance of the universe. We will account in this work for the radiation accumulated by star formation processes through most of the life of the universe, plus a contribution from active galactic nuclei (AGNs) to this wavelength range, known as the diffuse extragalactic background light (EBL).

The direct measurement of the EBL is a very difficult task subject to high uncertainties. This is mainly due to the contribution of zodiacal light, some orders of magnitude larger than the EBL (\eg \citealt{hauser01}; \citealt{chary10}). There are some measurements in the optical (\citealt{bernstein07}) and in the near-IR (\eg \citealt{cambresy01}; \citealt{matsumoto05}), but there is not general agreement about the reliability of these data sets (\citealt{mattila06}). In addition, these near-IR data appear to give intensity levels for the EBL in contradiction with the observation of very high energy (VHE, 30~GeV-30~TeV) photons from extragalactic sources (\citealt{aharonian06}; \citealt{mazin07}; \citealt{albert08}). Little is known about the mid-IR from direct detection due to the higher contamination from zodiacal light at those wavelengths. Measurements with the Far-Infrared Absolute Spectrometer (FIRAS) instrument on board the Cosmic Background Explorer, in the far-IR (\citealt{hauser98}; \citealt{lagache00b}), are thought to be more reliable. Other observational approaches set reliable lower limits on the EBL, such as measuring the integrated light from discrete extragalactic sources (\eg \citealt{madau00}; \citealt{fazio04}).

There are also other authors that focus on studying galaxy properties based on EBL results (\citealt{fardal07}), or on modeling a region of the EBL spectrum (\citealt{younger10}). On the other hand, there are phenomenological approaches in the literature that predict an overall EBL model (\ie between 0.1-1000~$\mu$m and for any redshift). These are basically of four kinds:
\begin{enumerate}
\item
Forward evolution, which begins with cosmological initial conditions and follows
a forward evolution with time by means of semi-analytical models (SAMs) of galaxy formation, \eg \citet{primack99}, \citet{somerville10} (hereafter, SGPD10) and \citet{gilmore10b} (hereafter, GSPD10).

\item
Backward evolution, which begins with existing galaxy populations and
extrapolates them backwards in time, \eg \citet{malkan98}, \citet{stecker06}, \citet{franceschini08} (hereafter, FRV08).

\item
Evolution of the galaxy populations that is inferred over a range of
redshifts. The galaxy evolution is inferred here using some quantity derived from observations such as the star formation rate (SFR) density of the universe, \eg Kneiske et al.~(2002), \citet{finke10}, \citet{kneiske10}.

\item
Evolution of the galaxy populations that is directly observed over the range of redshifts that contribute significantly to the EBL. The present paper, which we term empirical, belongs in this category.
\end{enumerate}

The type (i) SGPD10 and GSPD10 models discuss the same galaxy formation SAM but in different contexts: SGPD10 contains details of the model used in calculating the bolometric luminosity history of the universe and comparison with data, and GSPD10 focuses on the derived EBL and $\gamma$-ray attenuation. The SGPD10-GSPD10 model is based on an updated version of the semi-analytic theoretical approach described in \citet{somerville08} from the growth of super-massive black holes and their host galaxies within the context of the hierarchical Lambda Cold Dark Matter ($\Lambda$CDM) cosmological framework.  This is based in part on \citet{somerville99}, \citet{somerville01}, and in the simulations summarized by \citet{hopkins08a}, and \citet{hopkins08b}. We consider that these types of models are complementary to the observational approach taken here.

We consider the type (ii) FRV08 model the most complete observationally-based work of those mentioned above. They base their EBL modeling on galaxy luminosity functions (LFs), quantities which are directly observed and well understood. FRV08 exploit a variety of data to build evolutionary schemes according to galaxy morphology. They account for the contribution from early, late-type galaxies and a starburst population to the EBL. They use observed near-IR LFs from the local universe to $z=1.4$ for describing the early and late-type galaxies. For the starburst population they use an optical and only local LF. Different prescriptions are used to extrapolate the evolution of the different morphological types to higher redshifts, and corrections to fit their results to other observational data are applied.

Type (iii) models are not directly based on galaxy data. Instead, they are built from some parametrization of the history of the SFR density. This is a quantity derived using several different methods, each of which have different and signitficant uncertainties and biases. The SFR density is combined with uncertain assumptions about the emitted galaxy spectral energy distribution (SED) evolution as well.

Our type (iv) EBL estimates (the first approach in this category) will be compared in detail with the type (i) forward evolution semi-analytical galaxy formation model by SGPD10 and GSPD10, and the type (ii) oservationally motivated model by FRV08. The other works mentioned are briefly compared with our EBL calculations in Sec.~\ref{discussion}.

Our aim in this paper is to develop an EBL model that is as observationally-based and realistic as possible, yet fully reproducible, including a quantitative study of the main uncertainties in the modeling that are directly due to the data. This constrains the range of the background intensity and its implications to $\gamma$-ray astronomy. One important application of the EBL for $\gamma$-ray astronomy is to recover the unattenuated spectra of extragalactic sources. Our goal is to measure the EBL with enough precision that the uncertainties due to the EBL modeling, in these recovered unattenuated spectra, are small compared with other effects such as systematic uncertainties in the $\gamma$-ray observations. Examples of this are discussed in Sec.~\ref{attenuation}.

\begin{figure}
\includegraphics[width=\columnwidth]{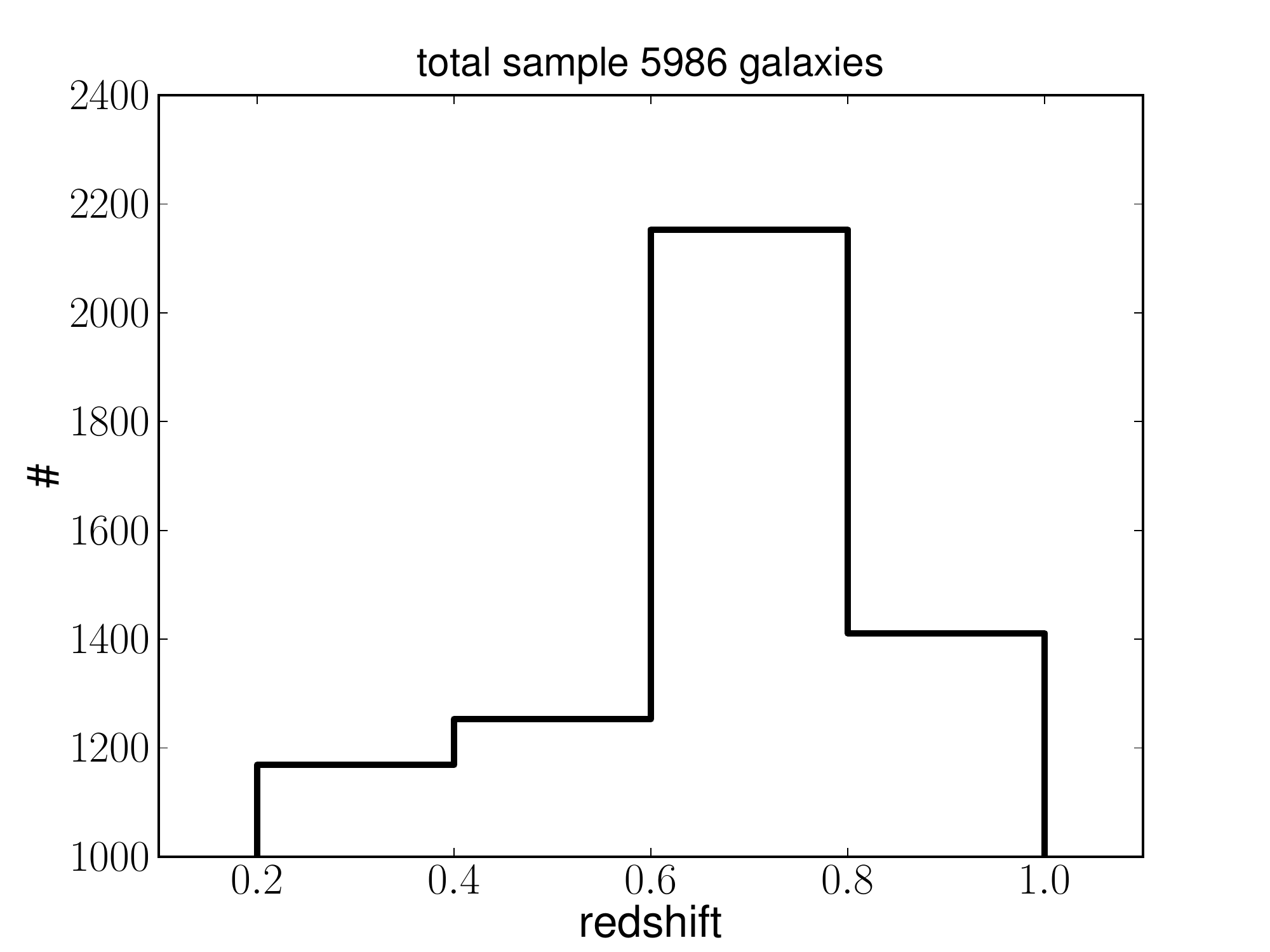}
\caption{Histogram of the number of galaxies versus redshift of our sample in the four redshift bins considered in this work.}
\label{fig1}
\end{figure}

Our model is based on the rest-frame $K$-band galaxy LF by \citet{cirasuolo10} (hereafter, C10) and on multiwavelength galaxy data from the All-wavelength Extended Groth Strip International Survey (AEGIS\footnote{http://aegis.ucolick.org/}, \citealt{davis07}; \citealt{newman10}) of about 6000 galaxies in the redshift range 0.2-1. These data sets are put together in a very transparent and consistent framework. The C10 LF is used to count galaxies (and therefore to normalize the total EBL intensity) at each redshift. The LF as well as our galaxy sample are divided in three magnitude bins according to the absolute rest-frame $K$-band magnitude \ie faint, middle, and bright (defined quantitatively later). Within every magnitude bin a SED-type is statistically attached to each galaxy in the LF assuming SED-type fractions that are function of redshift within those magnitude bins. This is estimated from fitting our AEGIS galaxy sample to the 25 galaxy SED templates from the SWIRE library. Then, luminosity densities are calculated from these magnitude bins from every galaxy population at all wavelengths, and finally all the light at all redshifts is added up to get the overall EBL spectrum. The results are linked with $\gamma$-ray astronomy and with the current understanding on galaxy evolution.

The paper is organized as follows: Section~\ref{data} describes the LF, our multiwavelength galaxy catalogue and the galaxy templates. Section~\ref{method} explains our methodology. The results for galaxy SED-type fractions, luminosity densities, SFR densities, EBL buildup, and EBL intensities are given in Section~\ref{results}. Section~\ref{attenuation} shows the attenuation computed from our EBL model for some VHE sources taken from the literature. In Section~\ref{discussion} our results are discussed including a detailed study on the uncertainties from the modeling, a comparison between our observationally-based EBL and that given by theoretical SAMs of galaxy formation. Finally, in Section \ref{summary} a summary with our main results and conclusions is presented.

Throughout this paper, a standard $\Lambda$CDM cosmology is assumed, with matter density $\Omega_{m}=0.3$, vacuum energy density $\Omega_{\Lambda}=0.7$, and Hubble constant $H_{0}=70$~km~s$^{-1}$Mpc$^{-1}$.

\section{Data description}
\label{data}

\subsection{$K$-band galaxy luminosity function}
\label{sec3.1}
The evolving galaxy LF in rest-frame $K$-band provided by C10 from $z=0$ to 4 is used. This evolving LF is the most accurate measurement to date of cosmological galaxy evolution in the near-IR, where dust absorption is less severe than in optical bands. The $k$-corrections in this band are less severe than in the optical as well. The choice of the C10 LF to normalize the model is also based on the smooth and well-studied shape of the galaxy SEDs in the near-IR, unlike others in UV or mid-IR wavelengths.

The resulting evolving LF is based on the UKIDSS Ultra Deep Survey (\citealt{lawrence07}), which has a large area and depth, and hence reduces the uncertainties due to cosmic variance and survey incompleteness. We refer the reader interested in details to that work. It is important to note that they give a parametrization of the evolution of the LF corrected from incompleteness and fitted by a Schechter function (\citealt{schechter76}) over redshift, $\Phi(M_{K}^{z},z)$, where $M_{K}^{z}$ is rest-frame $K$-band absolute magnitude at redshift $z$. The strongest assumption that they make is to keep constant the faint-end slope $\alpha$ in their parametrization.

\begin{table}
\centering
\begin{tabular}{|l|c|c|c|c|}
Band & $\lambda_{eff}$ [$\mu$m] & Observatory & Req. & UL [$\mu$Jy]\\
\hline
\hline
FUV & 0.1539 & GALEX & ext & -\\
NUV & 0.2316 & GALEX & ext & -\\
$B$ & 0.4389 & CFHT12K & det & -\\
$R$ & 0.6601 & CFHT12K & det & -\\
$I$ & 0.8133 & CFHT12K & det & -\\
$K_{S}$ & 2.14 & WIRC & det & -\\
IRAC~1 & 3.6 & IRAC & det & -\\
IRAC~2 & 4.5 & IRAC & obs & 1.2\\
IRAC~3 & 5.8 & IRAC & obs & 6.3\\
IRAC~4 & 8.0 & IRAC & obs & 6.9\\
MIPS~24 & 23.7 & MIPS & obs & 30\\
\hline
\end{tabular}
\caption{The photometric bands in our galaxy sample. For each we show the effective wavelength, the data source, the requirement for that band to be included for a given galaxy in our sample (det: a detection in this band is required; obs: observation in this band is required, but not necessarily a detection; ext: this band is considered extra information when available), and the 5$\sigma$ upper limit in that band in cases where there is no detection.}
\label{tab1}
\end{table}

\begin{figure*}
\includegraphics[width=12cm]{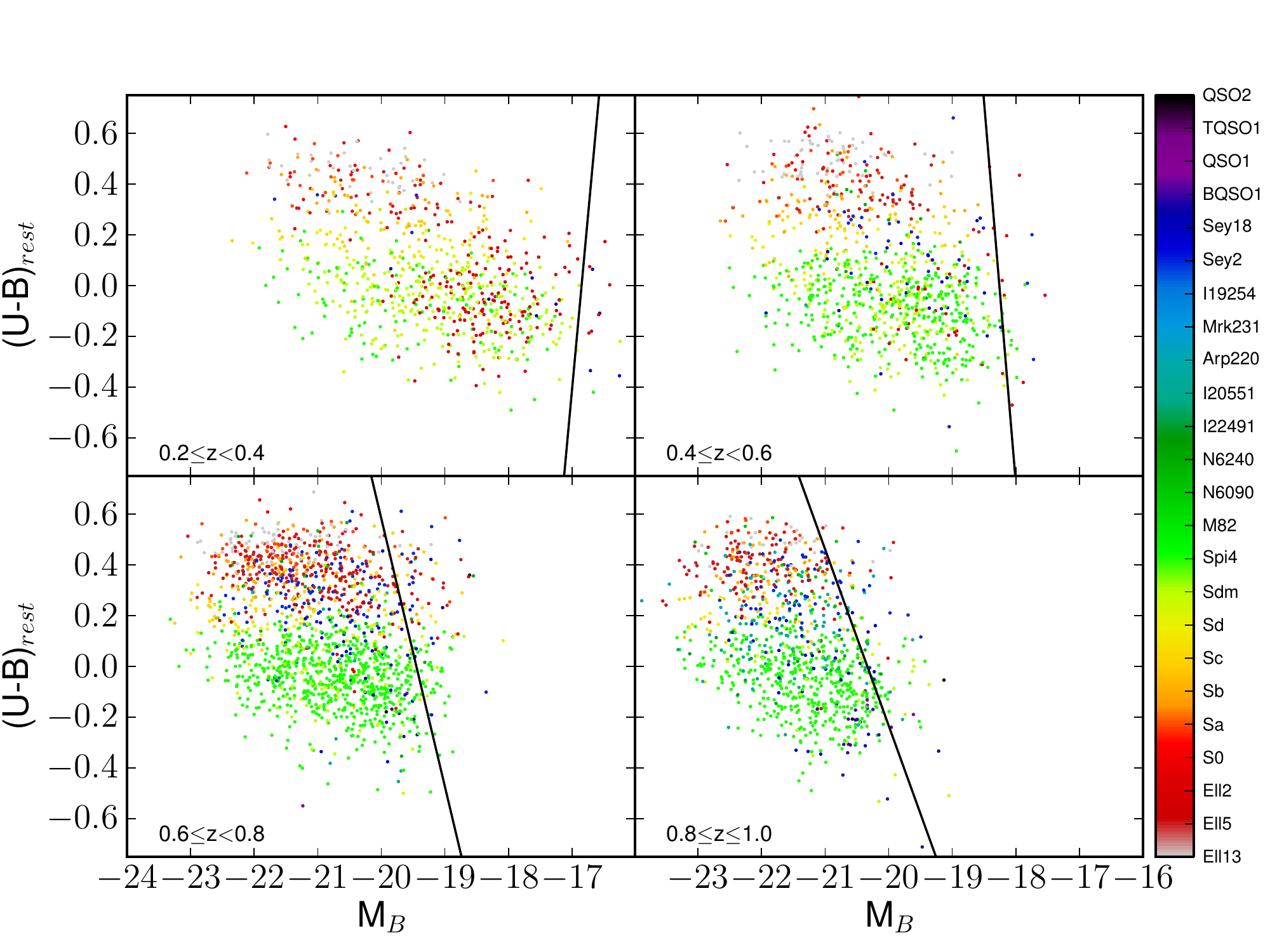}
\caption{Rest $U-B$ colour versus $B$-band absolute magnitude diagram for four different redshift bins to illustrate the incompleteness of our galaxy sample after the cuts explained in Sec.~\ref{sec4.1}. The black line is taken from Fig.~4 in \citet{willmer06}. Galaxies to the right of this line may suffer for a colour selection effect. The fractions of these galaxies are 1.8\%, 2.3\%, 7.3\% and 9.3\% for each of the redshift bins respectively. The colour code corresponds to the best-fitting galaxy SED-type from the SWIRE library (\eg Ell13, elliptical 13~Gyr old; Sa, early-type spiral; Spi4, very late-type spiral; I20551, starburst; Sey18, Seyfert galaxy 1.8, QSO2, quasi-stellar object with some ratio between optical and infrared fluxes). Magnitudes are in Vega system converted from AB system using the relations $U_{Vega}=U_{AB}-0.73$ and $B_{Vega}=B_{AB}+0.11$ from \citet{willmer06}.}
\label{fig2}

\end{figure*}
\begin{figure*}
\includegraphics[width=12cm]{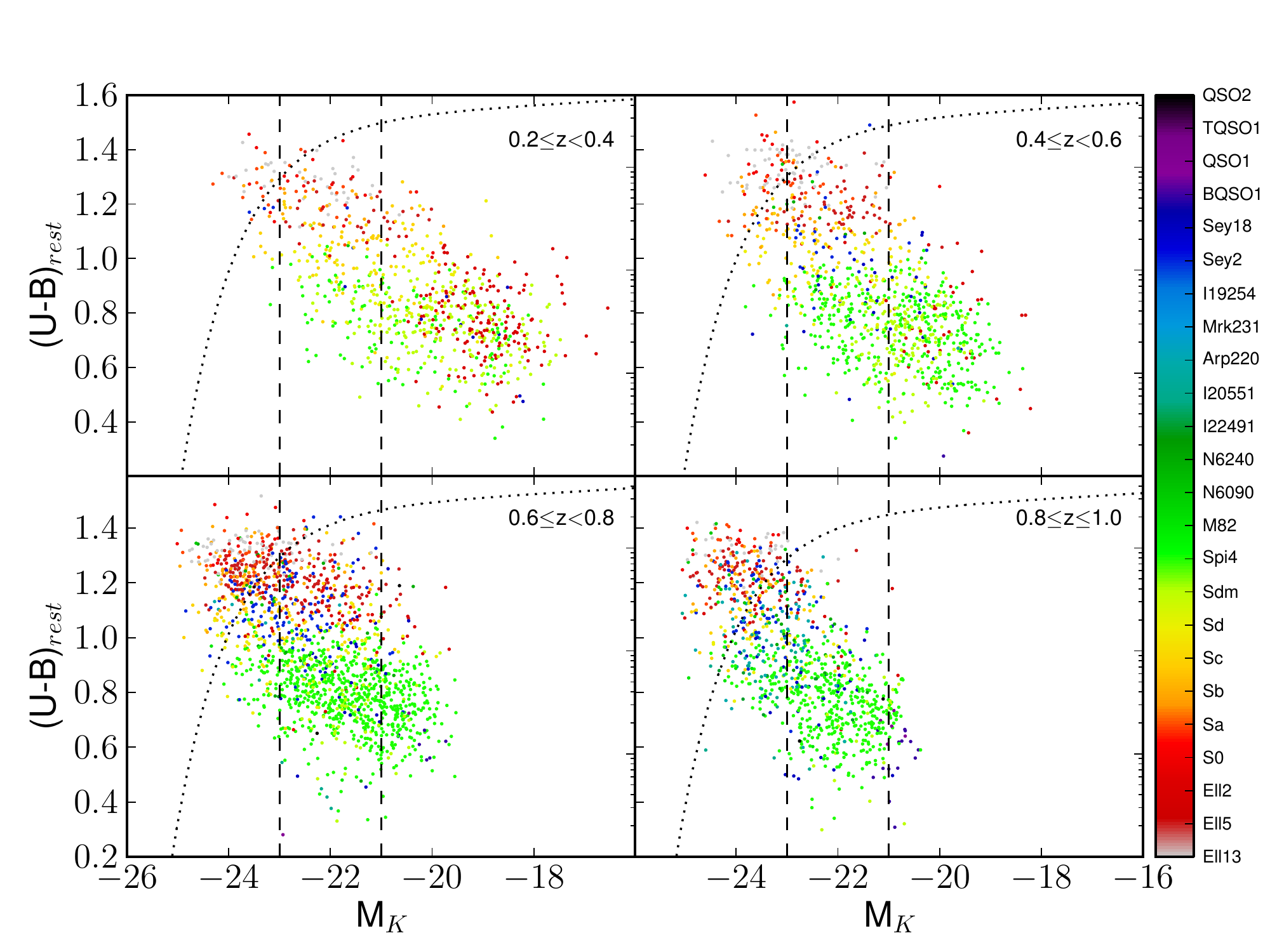}
\caption{Colour-magnitude diagram in the same four different redshift bins showing the galaxies of our sample after the cuts explained in Sec.~\ref{sec4.1}, for the magnitude bins defined in the text for the integrals in Eq.~\ref{jota}. It is over plotted the luminosity function by \citet{cirasuolo10} in the mean of every redshift bin with arbitrary units in the logarithmic $y$-axis. The colour code is the same that in Fig.~\ref{fig2}. Magnitudes are in AB magnitude system.}
\label{fig3}
\end{figure*}

\subsection{Galaxy sample description}
\label{sec3.2}
A multiwavelength galaxy catalogue built from AEGIS (\citealt{davis07}; \citealt{newman10}) for this work is used. This catalogue contains 5986 galaxies, all in the Extended Groth Strip (EGS). It is required that every galaxy in the sample have $5\sigma$ detections in the $B$, $R$, $I$, $K_{S}$ and IRAC~1 bands, and observations (but not necessarily detections) in the IRAC~2,~3,~4 and MIPS~24 bands (see Table~\ref{tab1}). These $5\sigma$ upper limits are given by the following fluxes: 1.2, 6.3, 6.9 and 30~$\mu$Jy, for IRAC~2, 3, 4 and MIPS~24 respectively, according to \citet{barmby08} for the IRAC bands and \citet{dickinson07} for MIPS~24. They are also summarized in Table~\ref{tab1}. In addition, 1129 of these galaxies have GALEX detections in the far-UV and 2345 galaxies in the near-UV. In our sample, 4376 galaxies have the highest quality spectroscopic redshifts measured by the Deep Evolutionary Exploratory Probe 2 team (DEEP2 DR3, \citealt{newman10}), with the DEIMOS spectrograph (\citealt{faber03}) on the Keck II telescope in an area of about 0.7~deg$^{2}$ in the sky. All the other galaxies in the sample (1610 galaxies) have secure photometric redshifts, more than 80\% with uncertainty in redshift less than 0.1. The redshift covered is between 0.2-1 (almost 60\% of the age of the universe) for a total sample of 5986 galaxies. For our purpose we will not distinguish between spectroscopic or photometric redshifts. This assumption will be discussed in Sec.~\ref{sec6.1}.

The optical photometry ($B$, $R$ and $I$ bands) was taken from imaging with the CFH12K camera (\citealt{cuillandre01}) on the Canada-France-Hawaii Telescope (CFHT) 3.6~m. The integration time for these observations was 1~hr in $B$ and $R$, and 2~hr in $I$, per pointing. More details may be found in \citet{coil04}.

The near-IR photometry in the $K_{S}$ band is from the Wide-field Infrared Camera (WIRC, \citealt{wilson03}) camera on the Hale 5~m telescope at the Palomar observatory. This data set is the most restrictive constraint on the area of our sample, therefore our galaxy catalogue is $K_{S}$ limited. The EGS field was surveyed to different depth for different sub-regions up to $K_{S}=22.5$ in the AB magnitude system. The details may be found in \citet{conselice08}.

The mid-IR data are from the IRAC and MIPS cameras on board the Spitzer Space Telescope. The details may be found in \citet{barmby08} and in \citet{dickinson07} describing the FIDEL survey, the source of our 24~$\mu$m data.

In addition, some data in the UV in two different bands 0.1530, 0.2310~$\mu$m from the Galaxy Evolution Explorer (GALEX, \citealt{morrissey07}) are included in our catalogue. This data set is part of the GALEX Deep Imaging Survey and the details may be found in \citet{salim09}.

Source catalogue from each of these imaging data sets where cross-matched using a Bayesian method, which took into account prior information from the surface densities of sources in each band (\citealt{huang10}). The IRAC~1 data were used as the primary reference catalogue.

It is important to note that all our data are public, except the MIPS~24 photometry, the cross-match catalogue, and the photometric-redshift catalogue (\citealt{huang10}). These will be released to the public soon.

The histogram of the redshift distribution of the AEGIS sample is shown in Fig.~\ref{fig1} in the four redshift bins considered in our calculations. Note the larger number around $z\sim0.7$, mainly due to the weighting scheme of the DEEP2 survey, which tends to select galaxies at $z>0.7$ based on colour-colour criteria, plus the effect in the opposite direction of losing faint galaxies at higher redshifts.

\begin{figure}
\includegraphics[width=\columnwidth]{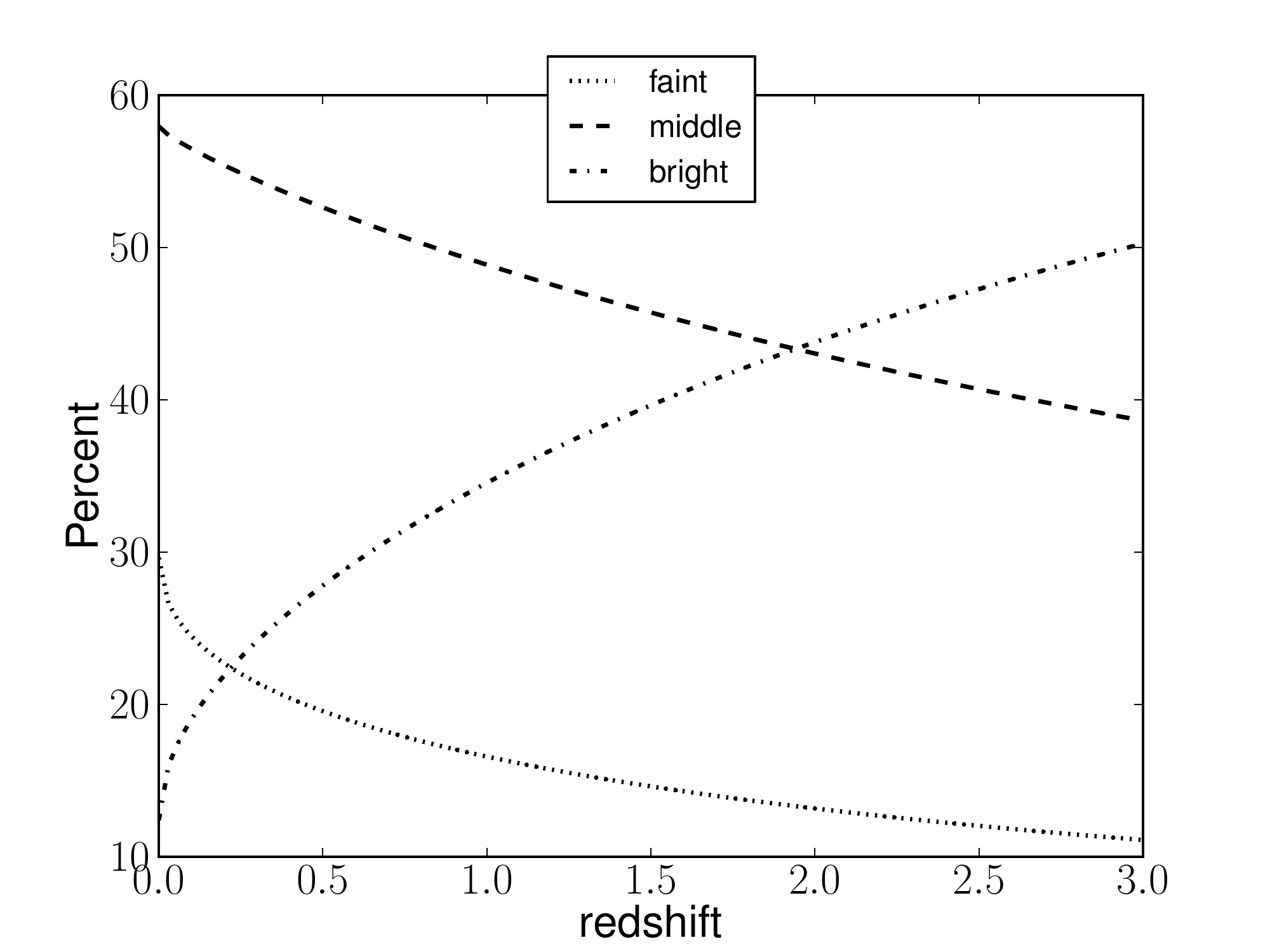}
\caption{Contribution from the three different magnitude bins defined in Sec.~\ref{method} to the total of the co-moving rest-frame $K$-band luminosity density calculated directly from the luminosity function (LF) by \citet{cirasuolo10}. The bulk of the light comes from the middle and bright-end of the LF, where the Schechter parameter $L_{\star}$ is. Note the increment with redshift of the bright-end contribution which decreases the impact of a possible color-selection effect or mis-typing (see Sec.~\ref{sec6.1}) at the highest redshift in our galaxy sample.}
\label{fig4}
\end{figure}

In order to calculate the absolute magnitudes in $U$, $B$ and $K$-band, we have computed the best-fitting template taken from the \citet{bruzual03} stellar population models to the data in the photometric bands $B$, $R$, $I$, $K_{S}$, IRAC~1 and FUV, NUV and IRAC~2 when available, using the code FAST (Fitting and Assessment of Synthetic Templates, see the Appendix in \citealt{kriek09} for details). FAST makes a $\chi^{2}$ minimization from a grid of \citet{bruzual03} models. We chose a stellar initial mass function (IMF) given by \citet{chabrier03}, an exponentially declining SFR $\sim \exp(-t/\tau)$ with $\tau$ ranging from 10$^{7}$-10$^{10}$~Gyr, (the same range for the ages), metallicities by mass fraction in the range 0.004-0.050 (solar metallicity is 0.02 in these units), and optical extinction $A_{v}$ from 0-10 following the \citet{calzetti00} extinction law. We calculate the absolute magnitudes from the best-fitting model using the $U$ Bessel filter, the $B$ filter from CFHT12K, and the same $K$-band filter from the UKIDSS survey, the same filter where the LF from C10 was estimated. All the transmission curves for these filters can be found in the default distribution of Le PHARE.

The sample was not corrected for incompleteness. However it is estimated here how this affects our results. The colour-dependent incompleteness of the DEEP2 survey was studied in \citet{willmer06}. They estimated a relation between the rest-frame $U-B$ colour versus the absolute magnitude in the $B$-band $M_{B}$ for which galaxies fainter than this relation have colour-dependent incompleteness. We show in Fig.~\ref{fig2} colour-magnitude diagrams of our AEGIS galaxy sample for four different redshift bins. The black line is taken from Fig.~4 in \citet{willmer06}. Galaxies located to the right of this line are likely missing. This figure is colour coded according to the calculated best-fitted SWIRE template (see Sec.~\ref{sec3.3} and Sec.~\ref{sec4.1}). The number of galaxies lying to the right of the relation, thus suffering colour-dependent incompleteness, are only of 1.8\%, 2.3\%, 7.3\% and 9.3\% for the different redshift bins presented in Fig.~\ref{fig2}.

Fig.~\ref{fig3} shows rest-frame $U-B$ versus absolute magnitude in the $K$-band in the three magnitude bins considered in this work to show an estimation on the galaxy number in each bin and their SED-types. We will describe this figure in the context of cosmological evolution in Sec.~\ref{sec6.3}.

Thus to recap, the normalizations of the EBL in our model is given by the $K$-band LF of C10, and our galaxy SED-type fractions give the relative contribution of every galaxy type to the total luminosity density. The assignment of SED-types to the galaxy population at a given redshift is done individually for three ranges in $K$-band absolute magnitude, as it will be discussed in Sec.~\ref{method}. Moreover, most of the contribution to the EBL (between 70-90\%) comes from around the knee of the LF ($L_{\star}$ according to the Schechter parametrization) as shown in Fig.~\ref{fig4} for the rest-frame $K$-band luminosity density (calculated directly from the integration of the C10 LF) and not from the faintest galaxy population where we suffer some small colour-dependent incompleteness. Fig~\ref{fig4} also shows that the contribution from the bright-end increases with redshift decreasing the impact of any colour-dependent effect. As the remaining colour-independent incompleteness does not have any effect on the galaxy SED-type fractions in our model or the overall normalization (which is set by the $K$-band LF), we conclude that our results are quite robust and the effect from incompleteness in our sample is minimal.

\subsection{Galaxy spectral energy distribution library}
\label{sec3.3}
\begin{figure}
\includegraphics[width=\columnwidth]{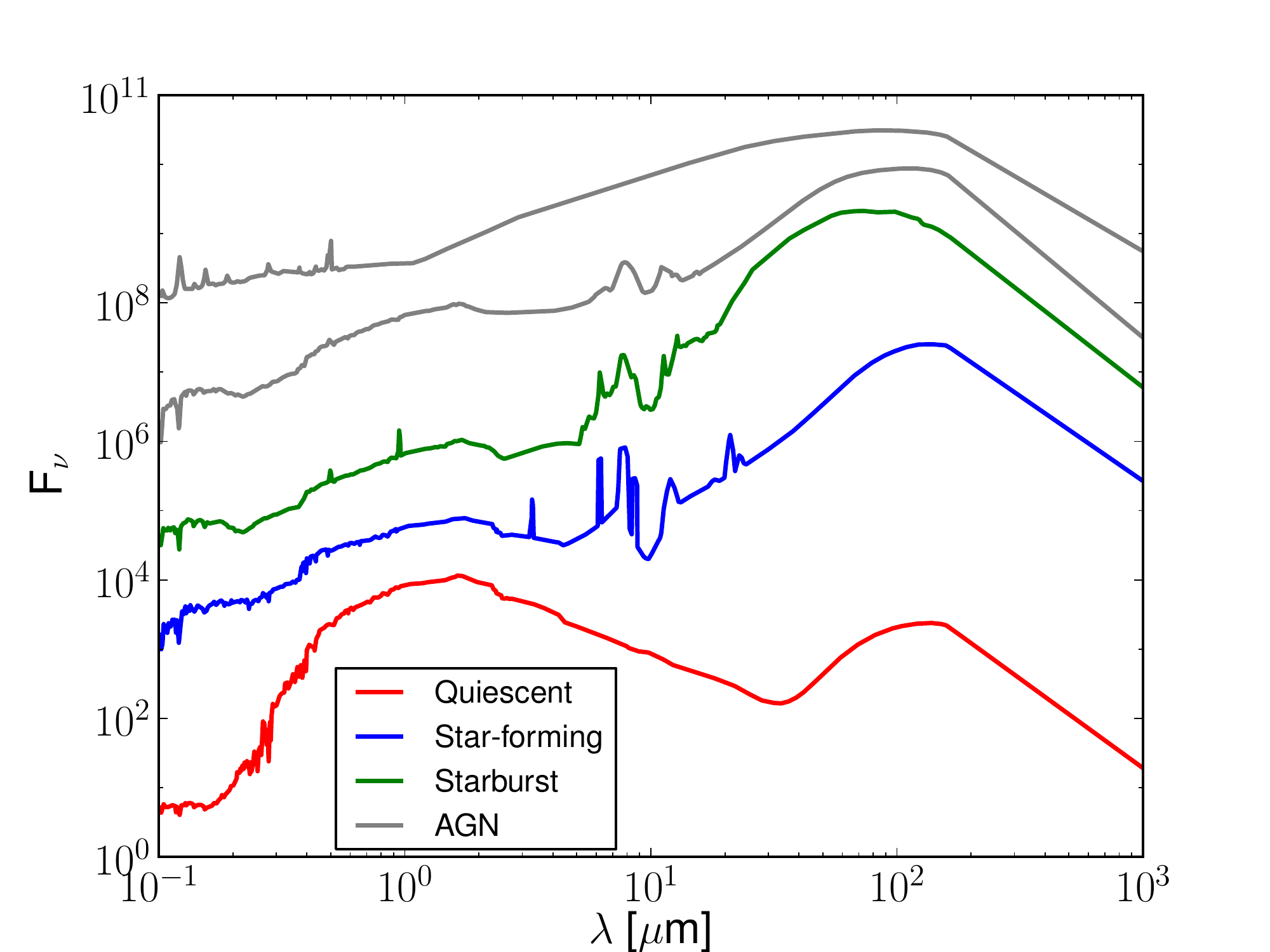}
\caption{Spectral energy distributions for some galaxy templates from the SWIRE library. We show here (from the bottom to the top) an early type quiescent galaxy (Ell13), a very late star-forming galaxy (Spi4), a starburst galaxy (I22491) and two different AGN galaxies: a Seyfert II, and a quasi-stellar object type I (QSO1). The $y$-axis is in arbitrary units.}
\label{fig5}
\end{figure}

The galaxy SEDs found in the SWIRE template library\footnote{http://www.iasf-milano.inaf.it/$\sim$polletta/templates/\\swire$\_$templates.html} (\citealt{polletta07}) are used. This library contains 25 templates, representative of the local galaxy population, defined as 3 elliptical galaxies, 7 spirals galaxies, 6 starbursts, 7 AGN galaxies (3 type I AGN, 4 type II AGN), and 2 composite (starburst+AGN) templates all covering the $\sim$~0.1-1000~$\mu$m wavelength range. The elliptical (quiescent), spiral (star-forming) and starburst (very star-forming) IR templates were generated with the GRASIL code (\citealt{silva07}) based on observations. The 7 spirals range from early to late types (\ie S0~-~Sdm). The starburst templates correspond to the SEDs of NGC~6090, NGC~6240, M~82, Arp~220, IRAS~22491-1808, and IRAS~20551-4250.  In all of the spiral and starburst templates, the spectral region between 5-12~$\mu$m, where many broad emission and absorption features are observed, was replaced by observed IR spectra from the PHT-S spectrometer on board the Infrared Space Observatory and from IRS on Spitzer. Some examples of these templates are shown in Fig.~\ref{fig5}.

We are aware that these templates do not include high-redshift galaxies ($z>0.3$). This effect will be taken into account in a future version of our EBL model when higher-redshift galaxy SEDs are released, including very vigorous starburts and AGNs not present in the local universe. The limitation of using local templates for high-redshift galaxies has been addressed in some works such as \citet{murphy09}, where they conclude that for ultra-luminous IR galaxies (ULIRGs) in the redshift range between $1.4<z<2.6$, IR luminosities are overpredicted when they are derived only using MIPS~24 photometry, thus showing a different behaviour than local ULIRGs. It has been also shown by \citet{menendez-delmestre09} that submillimeter galaxies in the redshift range between $0.65<z<3.2$ show a larger polycyclic-aromatic-hydrocarbon (PAH) emission than local analogs, suggesting a more extended dusty star-forming region than seen in local ULIRGS.

\section{Methodology}
\label{method}
The empirical approach of the EBL evolution directly observed over the range of redshifts that contribute significantly to the EBL is followed. This is type (iv) according to the classification given in Sec.~\ref{intro}. As briefly explained in Sec.~\ref{intro}, our aim is to calculate the EBL integrating over redshift luminosity densities. These quantities are estimated attaching statistically SEDs to the galaxies given by the LF by C10 in three different magnitude bins. This is achieved using galaxy SED-type fractions between $z=0.2-1$ by finding the best-fitting template of the 25 SED templates in the SWIRE library describing every galaxy in the AEGIS galaxy sample. Two different extrapolations for the galaxy SED-type fractions for $z>1$ are assumed leading to the same evolving EBL intensity from the UV to the mid-IR but different far-IR.

The Le PHARE v2.2 (Photometric Analysis for Redshift Estimations) code is used to find the best-fitting SWIRE SED template for every galaxy in the sample. Le PHARE is a publicly available code\footnote{http://www.cfht.hawaii.edu/$\sim$arnouts/LEPHARE/cfht\_lephare/} (\citealt{arnouts02}; \citealt{ilbert06}) mainly aimed to calculate photometric redshifts, but with the possibility to find the best-fitting template (among any library introduced as input) for galaxies with known redshift. Le PHARE makes use of a $\chi^{2}$ fitting procedure weighted from normalizations in every detected bands, and with the possibility to set upper limits for fluxes in some bands based on non-detections. From the fact that we have required observations in several bands to build our catalogue, we set for every galaxy in the fitting procedure upper limits on the bands where there is no a 5$\sigma$ detection. The information at all bands is used in the fitting.

\begin{figure*}
\includegraphics[width=18cm]{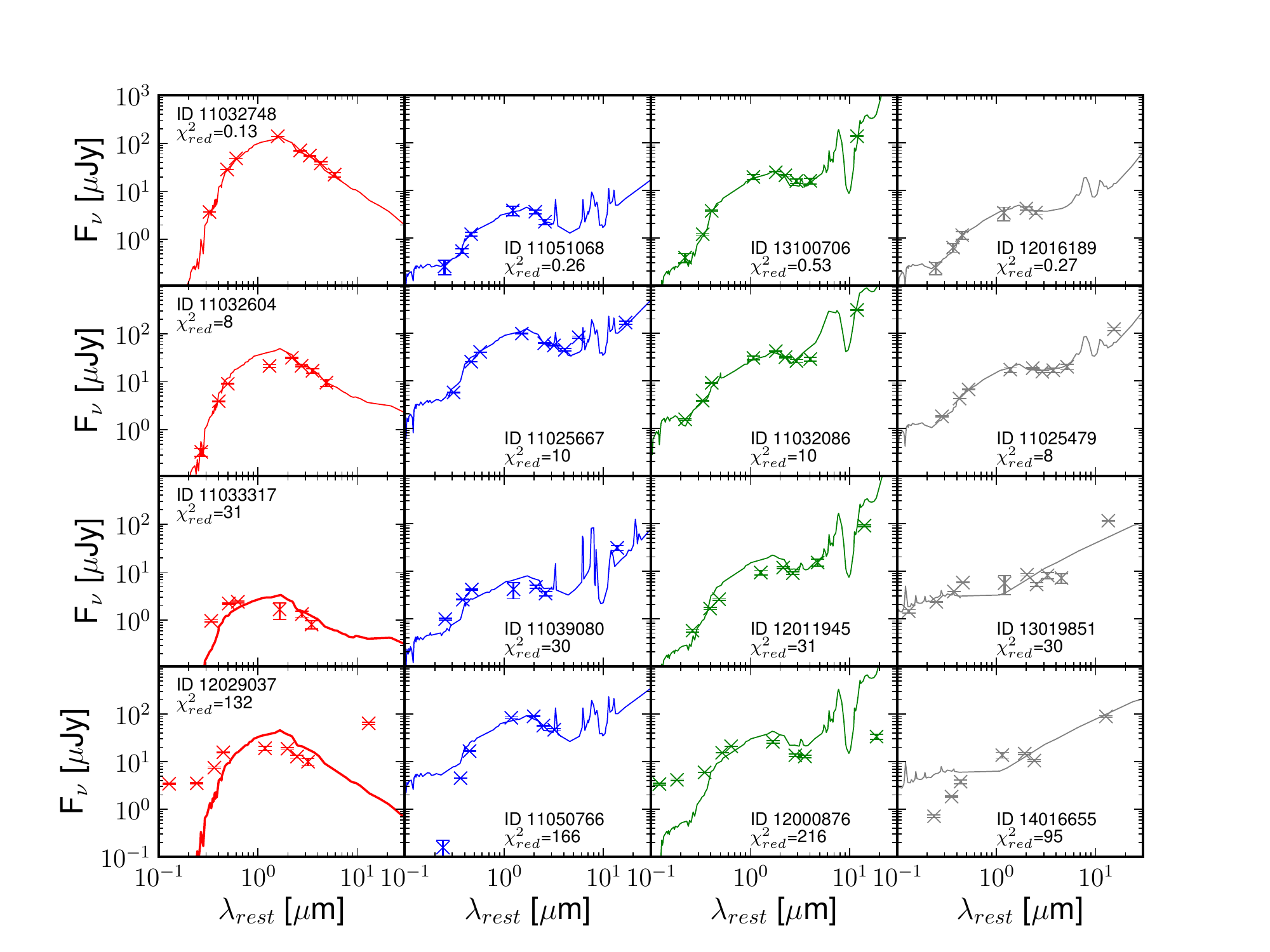}
\caption{Examples of the best fits (upper panel), fits around $\chi^{2}_{red}$=10 (second upper panel), fits around $\chi^{2}_{red}$=30 (second lower panel) and the worst fits (lower panel). The columns are from left to right: quiescent, star-forming galaxies, starbursts, and AGN galaxies. The AEGIS identification number is shown for the galaxy along with $\chi^{2}_{red}$ given by the fitting code Le PHARE described in Sec.~\ref{method}. The information at all bands is used in the fitting.}
\label{fig6}
\end{figure*}

For every galaxy, templates are rejected if they predict a flux in a given band that is higher than the upper limit for that band. The equations used for the fitting procedure are shown in Eq.~\ref{chi2}, with the parameter $s$ given by Eq.~\ref{s}:

\begin{equation}
\label{chi2}
\chi^{2}=\sum_{i} \Big(\frac{F_{obs,i}-sF_{temp,i}}{\sigma^{2}}\Big)^{2}
\end{equation}

\begin{equation}
\label{s}
s=\sum_{j} \Big(\frac{F_{obs,j}F_{temp,j}}{\sigma_{j}^{2}}\Big)/\sum_{j} \Big(\frac{F^{2}_{temp,j}}{\sigma_{j}^{2}}\Big)
\end{equation}

The notation is the following: $i$ refers to a given band, $j$ to the band used for the scaling, F$_{obs}$ is the observed flux, F$_{temp}$ is the flux from the templates, $\sigma$ is the 1$\sigma$ statistical uncertainty of the photometric measurement, and $s$ is the scaling factor that is chosen to minimize the $\chi^{2}$ values ($d\chi^{2}/ds=0$). For each galaxy in the sample, we use a redshift given by either spectroscopic or photometric data (see Sec.~\ref{sec3.2} for details on our sample).

We define the co-moving luminosity density at the rest-frame wavelength $\lambda$ as follows:

\[
j_{i}(\lambda,z)=j_{i}^{faint}+j_{i}^{mid}+j_{i}^{bright}=
\]
\[
\hspace{0.7cm}=\int_{M_{2}=-21.0}^{M_{1}=-16.6}\Phi(M_{K}^{z},z)f_{i}T_{i}(M_{K}^{z},\lambda)(1+z)dM_{K}^{z}+
\]
\[
\hspace{1.4cm}\int_{M_{3}=-23.0}^{M_{2}=-21.0}\Phi(M_{K}^{z},z)m_{i}T_{i}(M_{K}^{z},\lambda)(1+z)dM_{K}^{z}+
\]
\[
\hspace{2.1cm}\int_{M_{4}=-25.0}^{M_{3}=-23.0}\Phi(M_{K}^{z},z)b_{i}T_{i}(M_{K}^{z},\lambda)(1+z)dM_{K}^{z}
\]
\begin{equation}
\label{jota}
\hspace{4.9cm}\textrm{[erg~s$^{-1}$Mpc$^{-3}$Hz$^{-1}$]}
\end{equation}

\begin{table*}
\centering
\begin{tabular}{|c|c|c|c|c|c|}
$z_{mean}$ & Quiescent &  Star-forming & Starburst & AGN & Total non-rejected/rejected\\
\hline
\hline
0.3 & 235 (29\%)/77 (24\%) & 554 (69\%)/169 (52\%) & 1 (0\%)/23 (7\%) & 14 (2\%)/55 (17\%) & 804/324\\
0.5 & 157 (16\%)/38 (16\%) & 756 (77\%)/133 (47\%) & 13 (1\%)/13 (5\%) & 58 (6\%)/67 (32\%) & 984/241\\
0.7 & 328 (20\%)/59 (13\%) & 1079 (66\%)/149 (32\%) & 55 (3\%)/38 (8\%) & 175 (11\%)/221 (47\%) & 1637/467\\
0.9 & 144 (14\%)/22 (7\%) & 607 (58\%)/104 (32\%) & 164 (16\%)/21 (6\%) & 127 (12\%)/182 (55\%) & 1042/329\\
\hline
\end{tabular}
\caption{Galaxy SED-type fractions for our galaxy sample after applying the $\chi_{red}^{2}$ cuts (see Sec.~\ref{sec4.1}). Numbers are shown for galaxies non-rejected and rejected by the cuts, respectively. The total of non-rejected plus rejected galaxies is 5828. This is less than 5986, our total number of galaxies, because Le PHARE could not get any fit for 158 galaxies.}
\label{tab2}
\end{table*}

\noindent where $M_{K}^{z}$ in Eq.~\ref{jota} is the rest-frame absolute magnitude in $K$-band at redshift $z$. The SEDs are given by a function $T_{i}(M_{K}^{z},\lambda)=L_{\nu}$ (with units of erg~s$^{-1}$Hz$^{-1}$) with $i$ representing the different SWIRE SED types. We note that this function is dependent on $M_{K}^{z}$, since $T_{i}(M_{K}^{z},2.2)$ is the luminosity per Hz at the effective wavelength 2.2~$\mu$m for a galaxy with $M_{K}$. The fraction of faint, medium, and bright galaxies ($f_{i}$, $m_{i}$, $b_{i}$, the three different magnitude ranges) of each of the 25 classes is taken into account, according to their $M_{K}^{z}$, over redshift.

The magnitude ranges are defined in rest-frame $K$-band absolute magnitude using the AB magnitude system according to the different $M_{i}$ in the integration limits in Eq.~\ref{jota}. The faintest magnitude limit corresponds to the faintest galaxy in our sample, the medium range is chosen to lead to roughly equal numbers of galaxies in each bin in the three magnitude bins, and the brightest magnitude corresponds to the brightest galaxy in our sample. This separation is done to take into account the fact that for the same fraction of a given SED type, the contribution to the luminosity density will be different depending on luminosity. At any rate, fainter galaxies than $M_{1}$ are too faint to contribute significantly to the EBL even if their number density is fairly large. The same is true for galaxies brighter than $M_{4}$: in spite of their high luminosity their density is not high enough to contribute.

The function $\Phi(M_{K}^{z},z)$ in Eq.~\ref{jota} is the Schechter parametrization of the evolving LF as given by C10 in co-moving frame. The factor $(1+z)$ comes from the $k$-correction to account for the change in the definition of the local absolute magnitude $M_{K}^{0}$ with the redshift, \ie

\begin{equation}
\label{kcorrection}
k(z)=(1+z)\frac{T(M_{K}^{z},\lambda)}{T(M_{K}^{0},\lambda)}
\end{equation}

\begin{figure}
\includegraphics[width=\columnwidth]{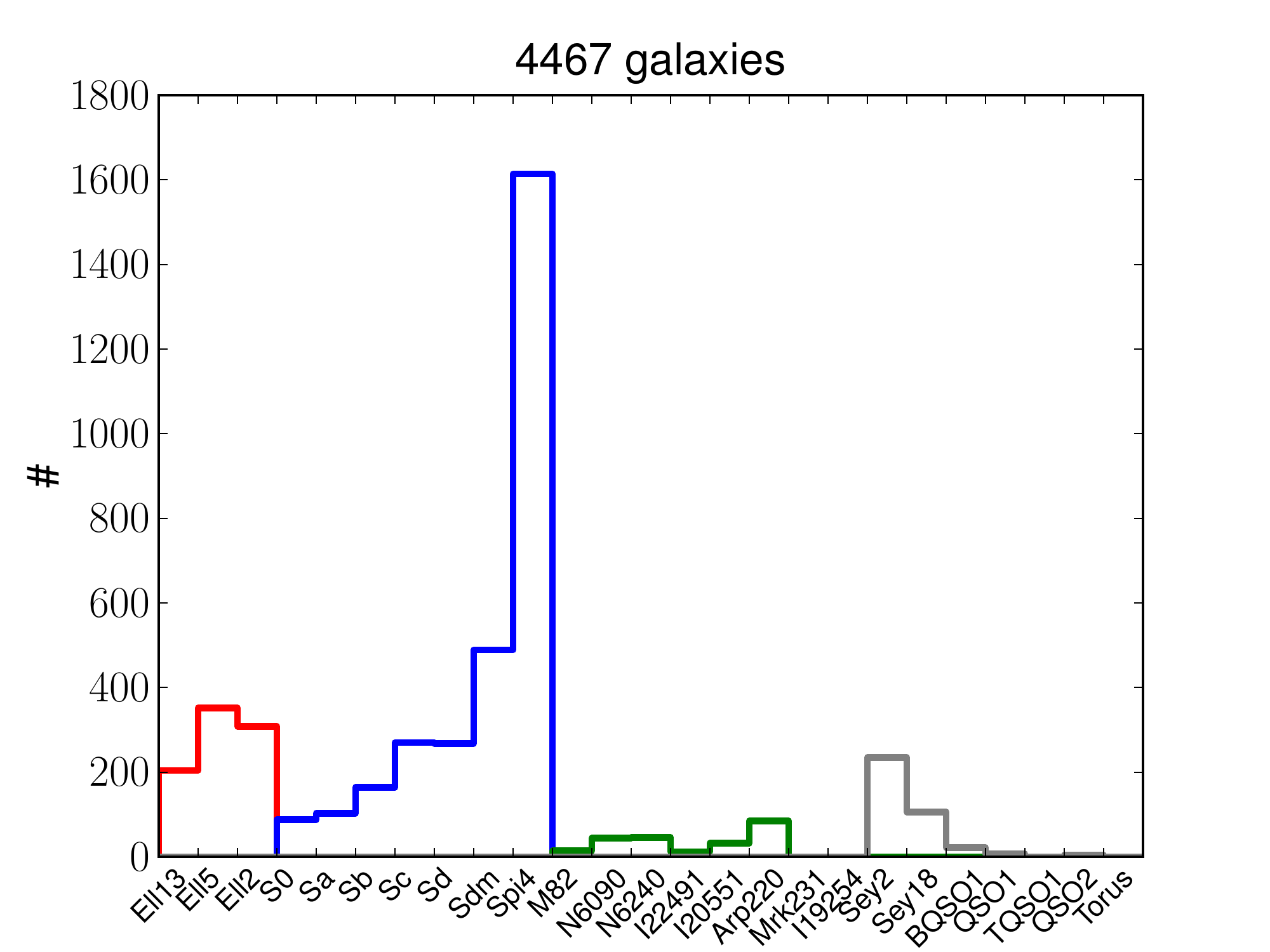}
\caption{Galaxy SED-types for the sample after removing the worst fits ($\sim25\%$ of the total sample, see Sec.~\ref{sec4.1}). We have 864 quiescent (in red, 19\%), 2996 star-forming galaxies (in blue, 67\%), 233 starbursts (in green, 5\%), and 374 AGN galaxies (in gray, 8\%) from a total of 4467 galaxies. The $x$-axis describes the names of the 25 SED templates from the SWIRE library.}
\label{fig7}
\end{figure}

The co-moving total luminosity density is calculated adding the luminosity density from the 25 SWIRE SED types, \ie
\begin{equation}
\label{jtotal}
j_{total}(\lambda,z)=\sum_{i}j_{i}(\lambda,z)
\end{equation}

We note that the total luminosity density at $2.2~\mu$m $j_{total}(2.2,z)$ is just the integral of the C10 LF.

\begin{figure}
\includegraphics[width=\columnwidth]{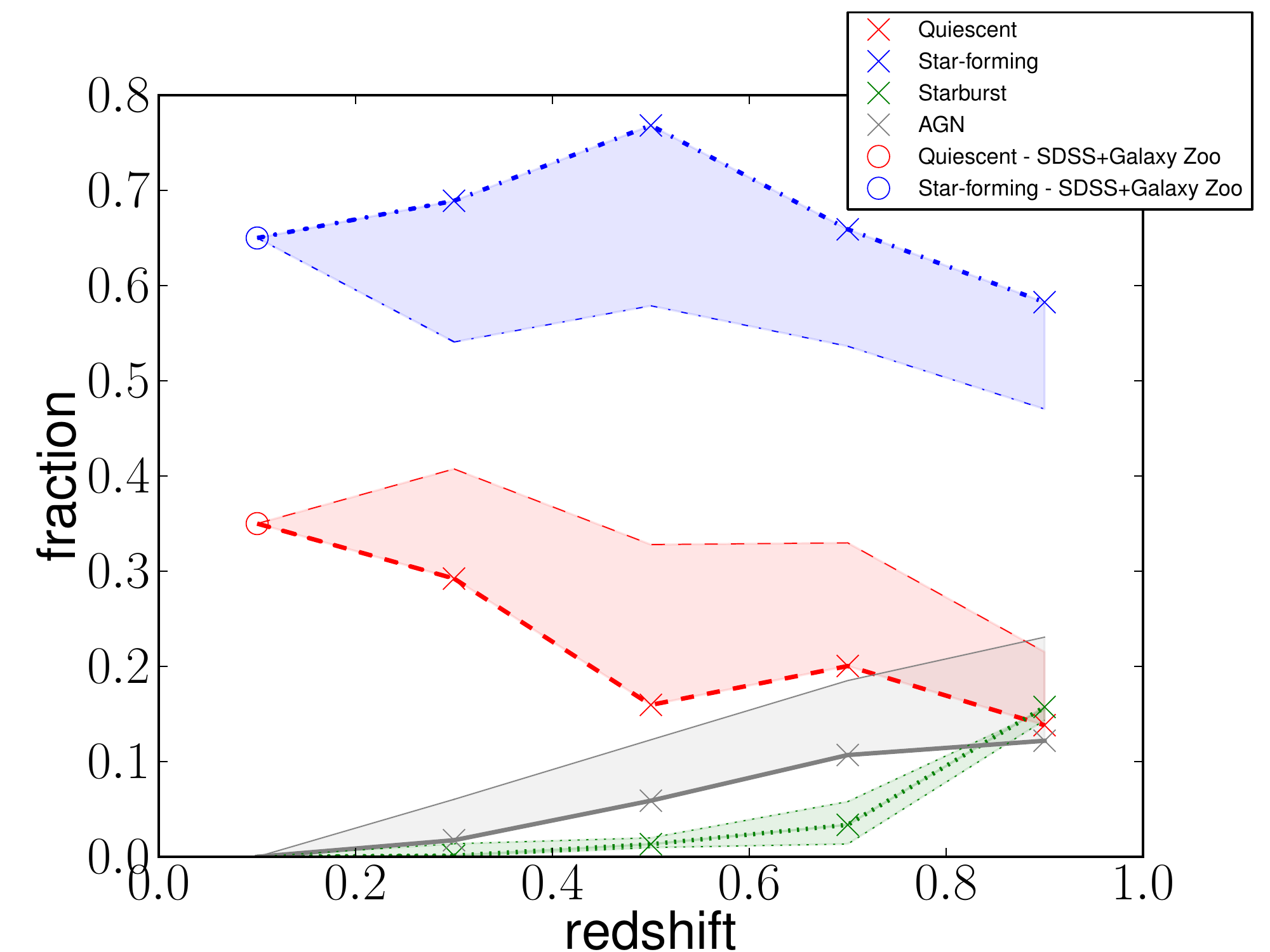}
\caption{Galaxy SED-type fractions from our catalogue (after the $\chi_{red}^{2}$ cuts, see Sec.~\ref{sec4.1}) of the different populations versus redshift according to our multiwavelength fits. We mark with crosses our fractions from $z=0.9-0.3$. The lines represent the linear interpolation that we use to calculate galaxy SED-type fractions for all redshift: dashed-red line represents quiescent galaxies, dotted-dashed-blue line represents star-forming galaxies, dotted-green line represents starburst galaxies, and solid-gray line represents AGN galaxies. The circles at $z=0.1$ are fractions computed from the SDSS-based sample (see text). We show with a shadow area the uncertainties from our lower limit for the errors as well as for our $\chi^{2}_{red}$ cut for fits. The uncertainties are around $\pm 0.1$.}
\label{fig8}
\end{figure}

The quantity defined by Eq.~\ref{jtotal} gives us an estimate of the total amount of light emitted by galaxies per unit volume at a given wavelength and redshift.

The history of the SFR density $\rho$ in the universe is then computed using the following approximation,
\begin{equation}
\label{star}
\rho=1.74\times10^{-10}(j_{IR}+3.3j_{2800})/L_{\odot}\hspace{0.3cm}\textrm{[M$_{\odot}$~yr$^{-1}$Mpc$^{-3}$]}
\end{equation} 
\noindent where $j_{IR}$ is the total bolometric infrared luminosity density integrated from 8-1000~$\mu$m, $j_{2800}$ is the luminosity density at 0.28~$\mu$m, and $L_{\odot}=3.839\times 10^{33}$~erg~s$^{-1}$ the solar bolometric luminosity. This equation is taken from \citet{wuyts09}, who add the UV and IR contributions (unobscured plus obscured), using calibrations for the local universe by \citet{kennicutt98} and a Salpeter IMF (\citealt{salpeter55}).

If Eq.~\ref{jtotal} is integrated from some redshift $z$ to $z_{max}=4$ (up to where the LF is given), the EBL flux seen by an observer at redshift $z$, due to the radiation emitted from $z_{max}$ down to $z$ is obtained,
\[
\lambda I_{\lambda}(\lambda,z)=\frac{c^{2}}{4\pi\lambda}\int_{z}^{z_{max}}j_{total}[\lambda (1+z)/(1+z'),z']\Big{|}\frac{dt}{dz'}\Big{|}dz'
\]
\begin{equation}
\label{ebl_spectrum}
\hspace{5.8cm}\textrm{[nW~m$^{-2}$sr$^{-1}$]}
\end{equation}

This is what we call co-moving EBL spectrum, which is given in intensity units. The factor $dt/dz'$ takes into account
the assumed cosmology (\eg \citealt{peebles93}), and is given explicitly by

\begin{equation} 
\label{peebles}
\Big|\frac{dt}{dz'}\Big|=\frac{1}{H_{0}(1+z')\sqrt{\Omega_{m}(1+z')^{3}+\Omega_{\Lambda}}}
\end{equation}
with $H_{0}$, $\Omega_{m}$ and $\Omega_{\Lambda}$ given by the parameters of the $\Lambda$CDM cosmology, exactly the same used by C10 for the LF.\\

In our approach, it is possible to directly calculate the contribution to the EBL from all redshift bins, as well as the evolution of the EBL spectrum with redshift and the processes related to this evolution, by sources of all the 25 SED types considered.

\section{Results}
\label{results}

\subsection{Galaxy SED-type fractions}
\label{sec4.1}

As explained in Sec.~\ref{method}, the Le PHARE code is used to fit every galaxy in our sample to the 25 SWIRE templates. For clarity, we will compress in our discussion (but not in our calculations, where they will remain independent) the 25 SED-types in the SWIRE library to four groups: quiescent, star-forming galaxies, starbursts, and AGN galaxies. We choose this nomenclature to clarify that our classification is multiwavelength-SED based, and not morphological.

We note that the fitting procedure is relatively sensitive to the errors on the photometric measurements leading to uncertainties in the galaxy SED-type fractions of $\pm 0.1$. For our model we set a lower limit of 6\% to all the photometric errors. The effect of different treatments of errors in the photometry is discussed thoroughly in Sec.~\ref{sec6.1} and it is shown in this section the uncertainties due to this effect on our galaxy SED-type fractions and on the EBL estimation.

\begin{table}
\centering
\begin{tabular}{|l|c|c|c|c|c|}
& $z=0.3$ & 0.5 & 0.7 & 0.9 & Total\\
\hline
\hline
faint & 507 & 411 & 251 & 49 & 1218\\
middle & 255 & 452 & 899 & 530 & 2136\\
bright & 43 & 121 & 487 & 462 & 1113\\
\hline
\end{tabular}
\caption{Number of galaxies (after applying the $\chi_{red}^{2}$ cuts, see Sec.~\ref{sec4.1}) in every magnitude and redshift bin used to calculate the galaxy SED-type fractions in Fig.~\ref{fig9}.}
\label{tab3}
\end{table}

To avoid accounting for bad fits, which do not correctly describe the galaxy photometric data, a cut in $\chi^{2}_{red}=\chi^{2}/n$ is applied, with $\chi^{2}$ given by Le PHARE (Eq.~\ref{chi2}) and $n$ degrees of freedom (bands with detections). We have checked carefully that $\chi^{2}_{red}\le30$ is a good value for quiescent, star-forming and starburst galaxies, but AGN galaxies are systematically worse fits, probably due to the fact that there is a large range in AGN SED shapes due to multiple emission components which cannot be easily encapsulated in a few templates.

Fig.~\ref{fig6} shows some examples of good and bad fits in for the four different main galaxy types with low $\chi^{2}_{red}$ in the top row, some fits around $\chi^{2}_{red}\sim10$ in the second row, other fits around $\chi^{2}_{red}\sim30$ the third row and some very bad fits (with very high $\chi^{2}_{red}$) in the bottom row. Due to the fact that AGN galaxies are systematically worse fits, two different cuts depending on the galaxy-SED type fitted are used for our model. These values are $\chi^{2}_{red}\le30$ for quiescent, star-forming and starburst galaxies, and $\chi^{2}_{red}\le10$ for AGNs. As for the uncertainties on the photometric errors, we show the uncertainties due to these cuts for the galaxy SED-type fractions and the EBL and discuss them in Sec.~\ref{sec6.1}.

After applying these cuts there are still 4467 galaxies remaining, \ie $\sim$75\% of the original sample. Fig.~\ref{fig7} shows a histogram of the galaxy SED-types in the total sample after the cuts, and the classification (shown with different colours) of the four main galaxy groups considered in this discussion. We find 19\% quiescent, 67\% star-forming galaxies, 5\% starbursts, and 8\% AGN galaxies.

\begin{figure}
\includegraphics[height=10cm,width=\columnwidth]{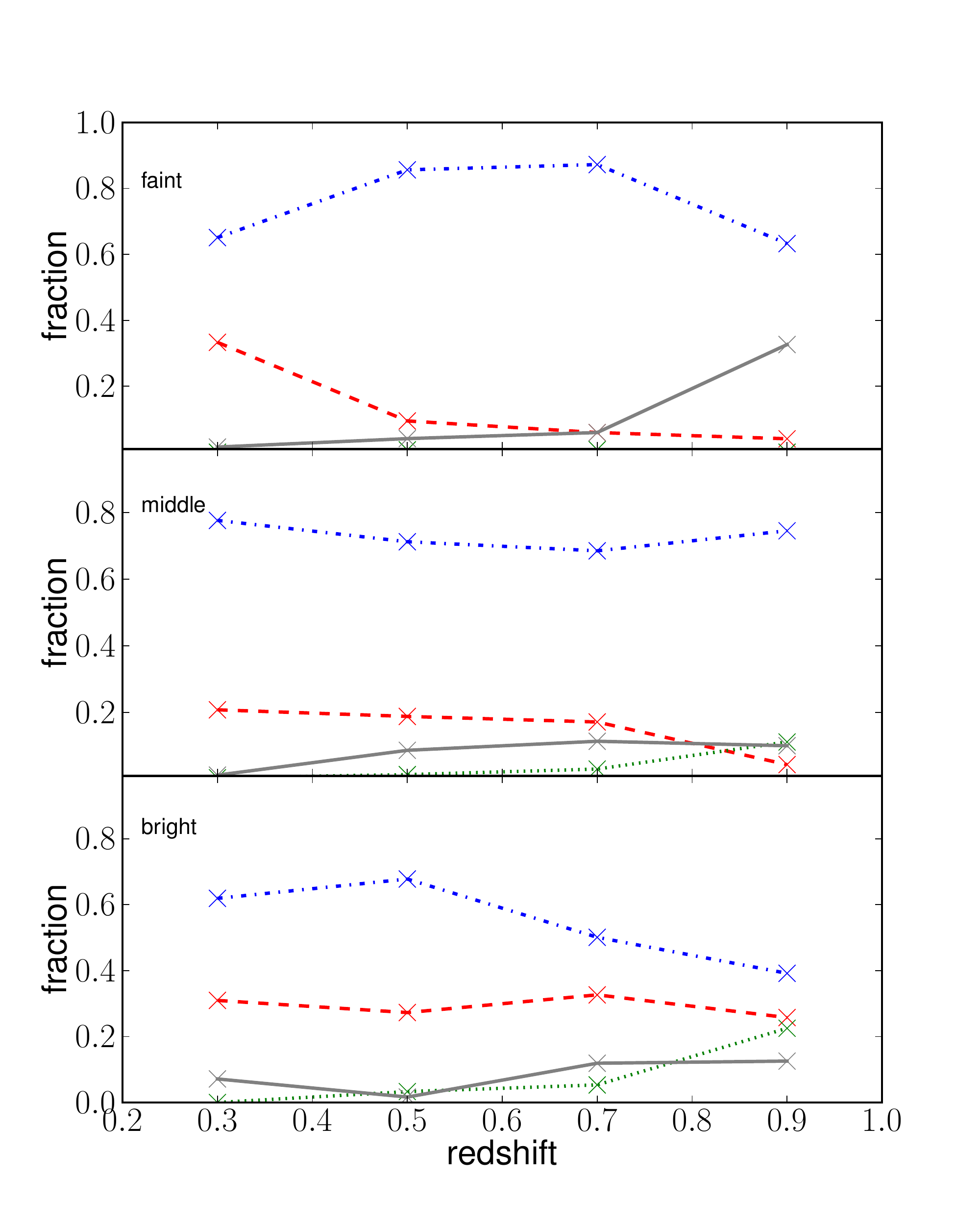}
\caption{Galaxy SED-type fractions (after applying the $\chi_{red}^{2}$ cuts, see Sec.~\ref{sec4.1}) of the different populations (the lines are the same as in Fig.~\ref{fig8}) versus redshift in the three different magnitude bins defined in the text for Eq.~\ref{jota}. See Table~\ref{tab3} for number of galaxies in every magnitude and redshift bin.}
\label{fig9}
\end{figure}

A bimodality between quiescent and star-forming galaxies is clearly found. Most of the quiescent galaxies are $\le 5$~Gyr old, late-type elliptical (Ell5 and Ell2, according to the SWIRE classification). The bulk of the star-forming population is late-type spirals with the PAH region measured using Spitzer data (Spi4, according to the SWIRE classification). In the starburst-like galaxies case, the Arp~220-like galaxies are dominant. The AGN-like population is clearly dominated by Seyfert-type galaxies, especially type II.

\begin{figure*}
\includegraphics[width=14cm]{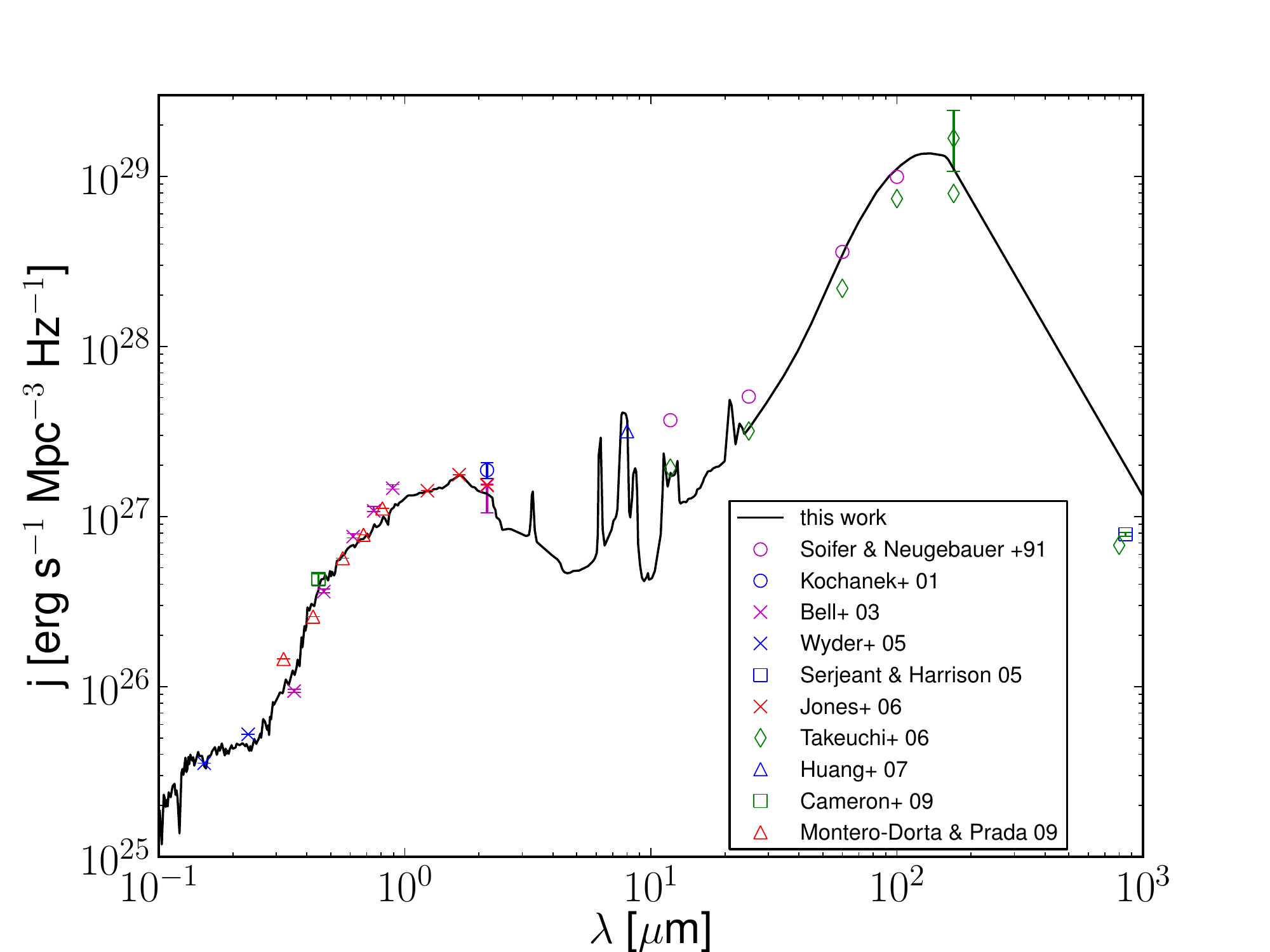}
\caption{Comparison between our estimation of the local luminosity density (black line) and observational data from different surveys: 12, 25, 60, 100~$\mu$m from \citet{soifer91}; $K$-band from \citet{kochanek01}; $u$, $g$, $r$, $i$, $z$, $K$-band from \citet{bell03}; $FUV$, $NUV$ from \citet{wyder05}; 850~$\mu$m from \citet{serjeant05}; $bj$, $rf$, $J$, $H$, $K$-band from \citet{jones06}; 12, 25, 60, 100, two different analysis for 170, 800~$\mu$m from \citet{takeuchi06} (two different analysis); 8~$\mu$m from \citet{huang07}; $B$-band from \citet{driver08} and \citet{cameron09}; and $u$, $g$, $r$, $i$, $z$ from \citet{montero-dorta09}.}
\label{fig10}
\end{figure*}

Table~\ref{tab2} and Fig.~\ref{fig8} show the galaxy SED-type fractions for four different redshift bins up to $z=1$, where we have chosen bins of $\Delta{z}$=0.2 for statistical reasons. The shadow regions are the uncertainties due to the lower limits on the photometric errors for the catalogue and for the $\chi^{2}_{red}$ cuts. This region is calculated changing the lower limits from 1-10\% in steps of 1\% and applying extreme cases for the cuts for every lower limit. The boundaries from these calculations lead to the shadow regions. The fractions adopted for the model are marked with crosses and wider lines. We observe that the fraction of quiescent galaxies increases by a factor $\sim 2$ from $z\sim0.9-0.3$, while the star-forming fraction keeps roughly constant for the full redshift range peaking at $z=0.5$. Starburst-type galaxies decrease very quickly from $z\sim0.9$ and reach almost 0 at $z\sim0.5$. On the other hand, the AGN-type fraction is roughly constant from $z\sim0.9-0.7$, and then decreases to 0.02 at $z\sim0.3$. This result should not be considered a complete picture of the evolution of the galaxy populations in the universe since these fractions depend on the color-magnitude limits of the survey as Fig.~\ref{fig2} shows. But what it is certainly described is the population of galaxies that contribute the most to the EBL around the knee of the LF (the middle and bright region of the LF, see Fig.~\ref{fig4}).

Fig.~\ref{fig9} shows the galaxy SED-type fractions for the three different magnitude bins defined in Eq.~\ref{jota} and explained in the previous section, Sec.~\ref{method}. These are the galaxy SED-type fractions used directly in Eq.~\ref{jota} to calculate the luminosity densities. Table~\ref{tab3} lists the number of galaxies in every magnitude and redshift bin used to estimate the galaxy SED-type fractions showed in Fig.~\ref{fig9}.

\begin{figure*}
\includegraphics[width=18cm]{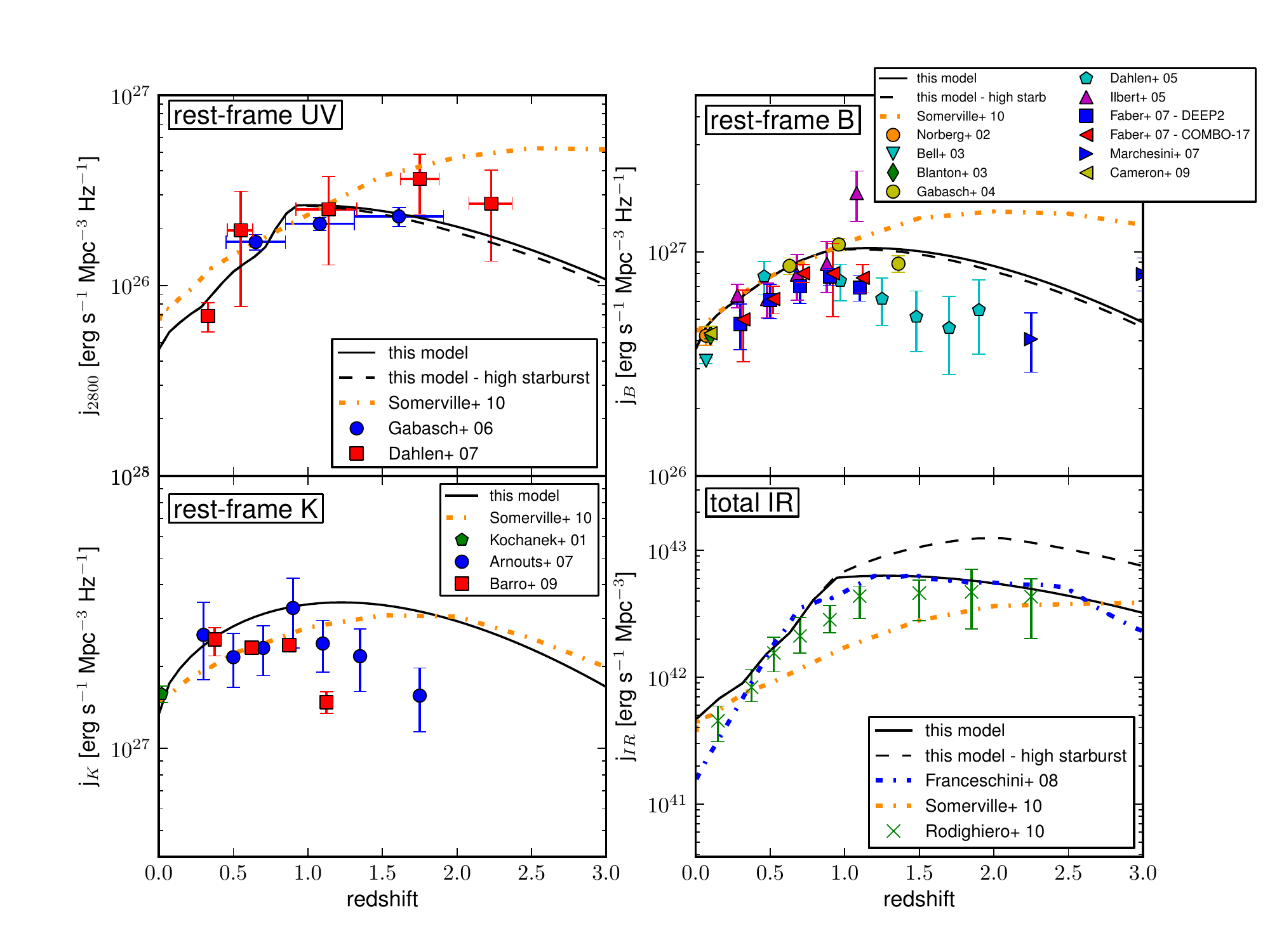}
\caption{Comparison between the calculated luminosity densities versus redshift for different spectral bands with observational data (solid-black line, for our \emph{fiducial} extrapolation; dashed-black line for our \emph{high-starburst} extrapolation for the galaxy SED-type fractions for $z>1$; see Sec.~\ref{sec4.1}). We also show as dot-dashed-orange line the model from \citealt{somerville10}. \emph{Upper-left panel}: rest-frame UV at 0.28~$\mu$m and data from \citet{gabasch06} and \citet{dahlen07}. \emph{Lower-left panel}: rest-frame $K$-band and observational data from \citet{arnouts07} and \citet{barro09}. It is important to note that this is just the integral of the LF by C10 between $M_{1}$ and $M_{4}$ in Eq.~\ref{jota}. \emph{Upper-right panel}: rest-frame $B$-band and observational data from a compilation from \citet{faber07} from these works: \citet{norberg02}, \citet{bell03}, \citet{blanton03a}, \citet{gabasch04}, \citet{dahlen05}, and \citet{ilbert05}. Data from \citet{marchesini07} and \citet{cameron09} are plotted as well. \emph{Lower-right panel}: integrated IR from 8-1000~$\mu$m data from \citet{rodighiero10} and the phenomenological estimations by Franceschini et al. (2008).}
\label{fig11}
\end{figure*}

The galaxy SED-type fractions are extrapolated to lower redshift ($z\sim 0$) by using results from \citet{goto03}, that use data from the Sloan Digital Sky Survey (SDSS). They give galaxy fractions according to a morphological classification. They are converted to SED classification by using two different observational works, using the Galaxy Zoo catalogue from SDSS data, on the abundances of blue-elliptical galaxies ($f_{be}=5.7\pm0.4\%$, \citealt{skibba09}) and red spirals ($f_{rs}\sim 25\%$ \citealt{schawinski09}). Utilizing these works we calculate galaxy SED-type fractions for the local $0<z<0.2$ universe as follows: in Fig.~12 and 15 of \citet{goto03} we see morphology-density and morphology-radius relations respectively. From the bin with the largest number density in any of those figures, we have the fractions of galaxies with early ($\sim 14\%$), intermediate-type ($\sim 26\%$), early-disc ($\sim 35\%$), and late-disc ($\sim 25\%$) morphology. The fractions of ellipticals are the fractions of early galaxies, $f_{ell}\sim 14\%$ and the fraction of spirals are the intermediate-type, plus the early-disc, plus late-disc galaxies, $f_{spi}\sim 86\%$. From the Galaxy Zoo papers (\citealt{skibba09}; \citealt{schawinski09}) these fractions are estimated for the local universe according to Eq.~\ref{eq9} and Eq.~\ref{eq10}: $f_{quies}\sim 35\%$ of quiescent and $f_{sf}\sim 65\%$ of star-forming galaxies.

\[
f_{quies}=f_{ell}-(f_{ell}\times f_{be})+(f_{spi}\times f_{rs})=
\]
\begin{equation}
\label{eq9}
\hspace{0.7cm}=0.14-(0.14\times 0.057)+(0.86\times 0.25)=0.35
\end{equation}

\[
f_{sf}=f_{spi}-(f_{spi}\times f_{rs})+(f_{ell}\times f_{be})
\]
\begin{equation}
\label{eq10}
\hspace{0.7cm}=0.86-(0.86\times 0.25)+(0.14\times 0.057)=0.65
\end{equation}

We have to keep in mind that these numbers are calculated from a different sample and a direct comparison with our sample may be not accurate. Note as well that our definition for quiescent and star-forming is not exactly the same as that the red and blue classification from Galaxy Zoo, but very similar. Some of our very early-type star-forming galaxies are red according to that classification, but the results do not change much because of the fewer number of these galaxies. In the opposite direction to this effect we note as well that Le PHARE prefers to fit some early-type star-forming galaxies as late-type red galaxies due to their bluer optical colours but very little dust re-emission, if any, according to the SWIRE templates.

To be able to compute the local EBL with accuracy, as well as its evolution out to the redshifts of the most distant objects detected by ground-based VHE $\gamma$-ray telescopes, \ie $z\le0.6$ (\citealt{albert08}), we would need to extrapolate the galaxy SED-type fractions to $z>1$. It is expected that the local EBL has contributions from these larger redshifts, although the behavior is different for the optical/near-IR and the far-IR due to the spectral region where the different populations contribute.

For the high-redshift universe ($z>1$, where there are no galaxies in our sample) two different cases are considered for the evolution of the galaxy SED-type fraction. It is shown that our results are not changed significantly except in the far-IR by these two choices. For the redshifts less than those of the most distant known $\gamma$-ray sources, and redshifts where future sources are likely to be found in the near future by Imaging Atmospheric Cherenkov Telescopes (IACTs), it is found almost no change in the EBL even with a fairly large adjustment in the evolution of galaxy-SED type fractions. This is discussed in Sec.~\ref{sec6.1} and here we show the uncertainties of the EBL and others quantities calculated due to these assumptions. The \emph{fiducial} choice is to keep constant the fractions computed for our highest redshift bin. This choice is made for simplicity, due to the difficulty in the multiwavelength classification of distant galaxies with current instruments. But we do note that there is strong evidence from several observational results by \citet{reddy05}, \citet{perezgonzalez08}, \citet{taylor09} and \citet{wuyts09} which suggest no further evolution at higher redshifts of the quiescent population. All these independent works claim that the fraction of distant non star-forming red objects in the high-redshift universe keep constant around 24-33\% of the total number of galaxies up to $z>2.5$. We find at $z\sim0.9$ around 14\% of quiescent galaxies, which it is kept constant for higher redshifts. We note here the red-galaxy incompleteness for DEEP2, implying that our fractions might underestimate the actual number of mainly quiescent galaxies in the faint-end of the LF (as seen in the very low number of quiescent fractions in Fig.~\ref{fig9}), due to the difficulty for the DEEP2 survey to characterize faint-red galaxies for $z>0.8$. The impact of this effect is decreased by taking into our catalogue galaxies with photometric redshift. In any case, there are no consequences for the EBL results as previously discussed in Sec.~\ref{sec3.2}, because the bulk of the light comes from the region of the LF around $L_{\star}$ where we are basically complete.

As alternative approach, we choose to increase linearly with redshift the starburst-like fraction from our calculated 16\% at $z=0.9$ up to 60\% at $z=2$, while decreasing at the same rate the quiescent and star-forming galaxies. The weight of every one of the 25 SWIRE templates is changed in the same proportion. The fractions are kept constant at $z=2$ for $z>2$. This approach is called \emph{high-starburst} and it is used to determine a likely upper limit on the EBL at long wavelengths (see Sec.~\ref{sec4.3}).

\subsection{Luminosity densities}
The local galaxy luminosity density is shown in Fig.~\ref{fig10} calculated using Eq.~\ref{jtotal}. The solid-black line is computed from the sum of the contributions from all the 25 SED-types. An excellent agreement is found with the observational data from independent surveys over all wavelengths. We note that our different assumptions for the high-redshift fractions lead necessarily to the same result because this is the light emitted at $z=0$.

Fig.~\ref{fig11} shows the evolution over redshift of the luminosity densities at different wavelengths for both of our extrapolations for the high-redshift fractions. We show with dot-dashed-orange line the galaxy formation SAM prediction by SGPD10 and postpone the discussion to Sec.~\ref{SAMs}. The upper-left panel shows the rest-frame 0.28~$\mu$m, which is in good agreement with the observational results by \citet{gabasch06} and \citet{dahlen07} for $z<1.5$, and somewhat lower between $z=2-1.5$ than the \citet{dahlen07} data. This quantity is also directly related with the SFR density through Eq.~\ref{star}. The upper-right panel shows the rest-frame $B$-band, which is in good agreement with some observational results, such as \citet{norberg02}, \citet{gabasch04}, \citet{ilbert05}; but around a 15-20\% higher than \citet{faber07}. At $z>1$ we are a factor $\sim2$ higher than the data by \citet{dahlen05}. The lower-left panel shows the rest-frame $K$-band, among some observational results by \citet{arnouts07} and \citet{barro09}. The lower-right panel shows the evolution of our calculated total IR luminosity over redshift, that given by the FRV08 model and from observations by \citet{rodighiero10}. We note a general good agreement with these data (we are a a factor 1.5 higher around $z\sim1$) for our \emph{fiducial} extrapolation of SED-types beyond $z=1$, but we are predicting a higher luminosity density for the \emph{high-starburst} assumption. The agreement with FRV08 is pretty good, except for the lowest redshifts. The total IR luminosity is also directly related with the SFR density through Eq.~\ref{star}.

\begin{figure}
\includegraphics[width=\columnwidth]{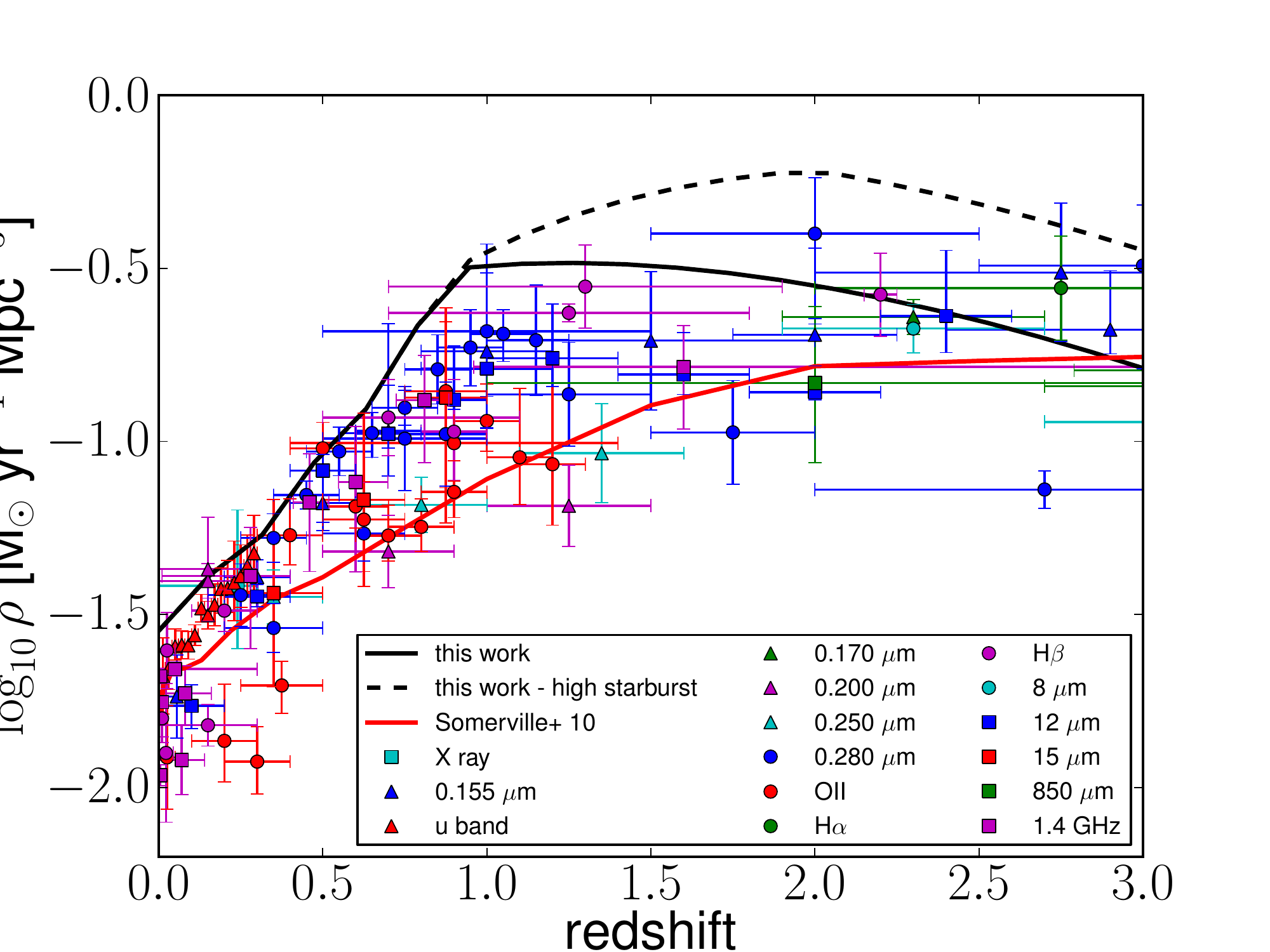}
\caption{Comparison between the calculated star formation rate density computed using Eq.~\ref{star} for a Salpeter initial mass function, the prediction using the same equation from the luminosity densities provided by \citet{somerville10} (red-solid line), and some observational data from different estimators shown in the legend. The compilation of data points is taken from \citet{perezgonzalez08}. The solid and dashed-black lines are from the different extrapolations for the galaxy SED-type fractions for $z>1$ (see Sec.~\ref{sec4.1}).}
\label{fig12}
\end{figure}

\begin{figure*}
\includegraphics[width=18cm]{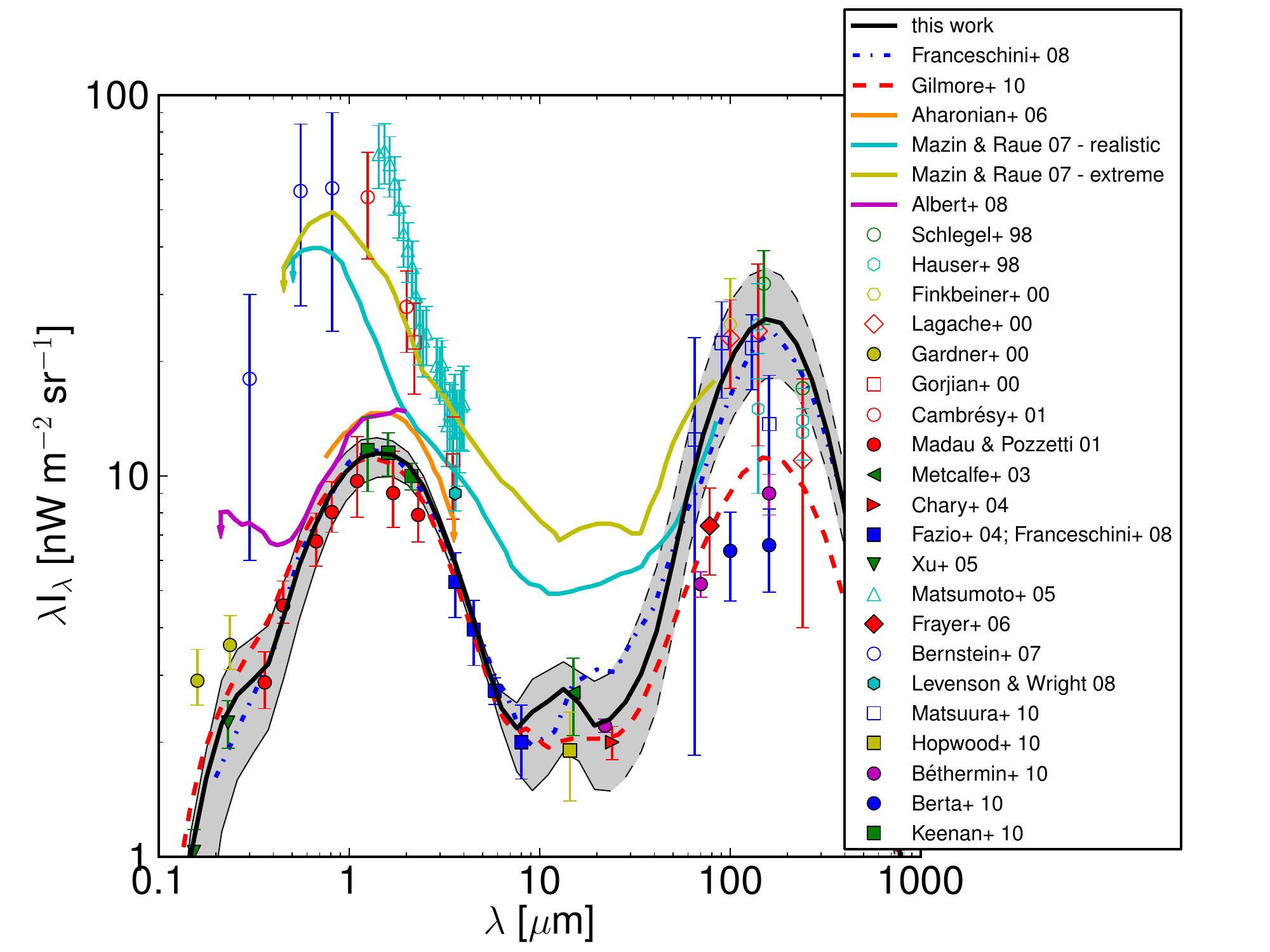}
\caption{The solid-black line is the extragalactic background light calculated by the \emph{fiducial} extrapolation of the galaxy SED-type fractions for $z>1$. Empty symbols are direct measurements: 0.3, 0.555, 0.814~$\mu$m by \citet{bernstein07}; 1.43, 1.53, 1.63, 1.73, 1.83, 1.93, 2.03, 2.14, 2.24, 2.34, 2.44, 2.54, 2.88, 2.98, 3.07, 3.17, 3.28, 3.38, 3.48, 3.58, 3.68, 3.78, 3.88, 3.98~$\mu$m by \citet{matsumoto05} using IRTS; 1.25, 2.2~$\mu$m (slightly shifted for clarity) by \citet{cambresy01}; 2.2, 2.5~$\mu$m by \citet{gorjian00}; 60, 100~$\mu$m by \citet{finkbeiner00} all these using DIRBE; 65, 90, 140 (slightly shifted for clarity), 160~$\mu$m by \citet{matsuura10} using AKARI; 100, 140, 240~$\mu$m by \citet{lagache00b}; 140 (slightly shifted for clarity), 240~$\mu$m by \citet{schlegel98}; 140, 240~$\mu$m by \citet{hauser98} all these using FIRAS. Filled symbols are galaxy-count data, usually considered lower limits: 0.1530, 0.2310~$\mu$m by \citet{xu05} using GALEX; 0.1595, 0.2365~$\mu$m by \citet{gardner00} using HST and STIS; 0.36, 0.45, 0.67, 0.81, 1.1, 1.6 (slightly shifted for clarity), 2.2~$\mu$m (slightly shifted for clarity) by \citet{madau00} using HST and ground-based telescopes; 1.25, 1.60, 2.12~$\mu$m by \citet{keenan10} using Subaru; 3.6~$\mu$m by \citet{levenson08}; 3.6, 4.5, 5.8, 8.0~$\mu$m by \citet{fazio04} with a reanalysis of the last point by Franceschini et al. 2008 all these using IRAC; 15~$\mu$m by \citet{metcalfe03} using ISO; 15~$\mu$m by \citet{hopwood10} using AKARI; 24~$\mu$m by \citet{papovich04} and \citet{chary04}; 24 (slightly shifted for clarity), 70, 160~$\mu$m by \citet{bethermin10} using MIPS; 71.4~$\mu$m by \citet{frayer06} using MIPS; 100, 160~$\mu$m by \citet{berta10} using Herschel. The coloured-solid lines (\citealt{aharonian06}; \citealt{mazin07}; \citealt{albert08}) are upper limits from $\gamma$-ray astronomy using different blazars (see Sec.~\ref{attenuation} for details). The dot-dashed-blue line, and the dashed-red line are the predictions from the models by Franceschini et al. (2008) and \citet{gilmore10b}, respectively. Uncertainties in the our EBL estimation are shown with a shadow area. These EBL uncertainties include the uncertainties in Schechter parameters of the LF by \citet{cirasuolo10}, photometric errors in the galaxy catalogue, $\chi^{2}_{red}$ cuts applied and extrapolations of the galaxy SED-type fractions for $z>1$ (see Sec.~\ref{sec4.1}). The envelope of the shadow region within the dashed line at wavelengths above 24~$\mu$m shows the region where there is no photometry in our galaxy catalogue. The EBL uncertainties are thoroughly discussed in Sec.~\ref{sec6.1}.}
\label{fig13}
\end{figure*}

\begin{figure*}
\includegraphics[width=18cm]{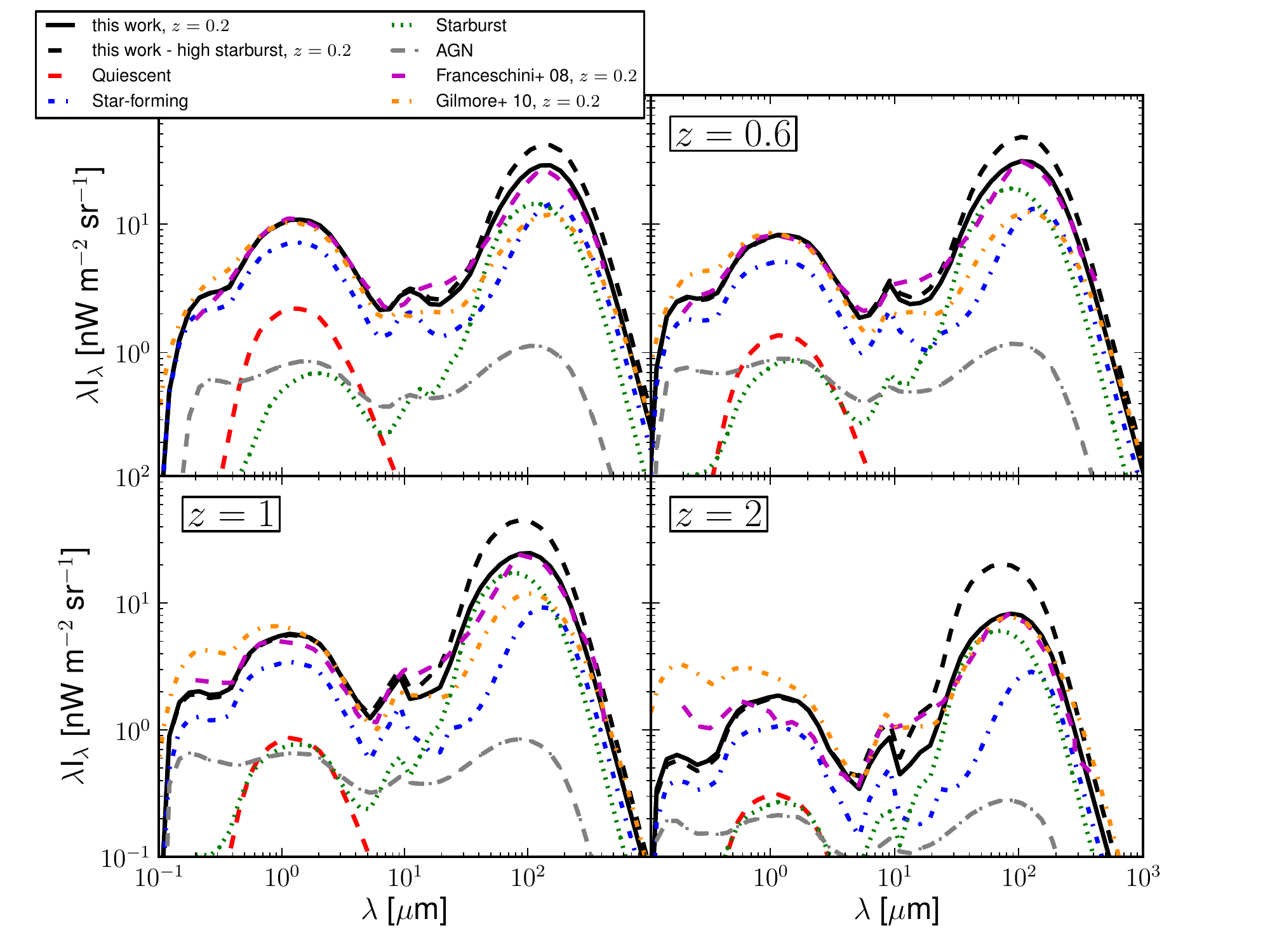}
\caption{Extragalactic background light (EBL) in co-moving frame predicted by our model at different redshifts for the two assumptions for the extrapolation of the fractions for $z>1$ (see Sec.~\ref{sec4.1}). The contribution to the EBL from quiescent (red-dashed line), star-forming galaxies (blue-dotted-dashed line), starbursts (green-dotted line), and AGN galaxies (gray-dotted-long-dashed line) to the \emph{fiducial} model are shown. For comparison, the predictions from other models are shown using magenta-dashed line for Franceschini et al. (2008) and orange-dot-dashed line for \citet{gilmore10b}.}
\label{fig14}
\end{figure*}

\begin{table*}
\centering
\begin{tabular}{|c|c|c|c|c|c|}
$z$ & Quiescent &  Star-forming & Starburst & AGN & $I_{bol}$ [nW~m$^{-2}$sr$^{-1}$]\\
\hline
\hline
0.0 & 4.71 (7\%) & 39.70 (57\%) & 20.45 (30\%) & 4.41 (6\%) & 69.26\\
0.2 & 3.86 (5\%) & 38.96 (54\%) & 24.54 (34\%) & 5.25 (7\%) & 72.60\\
0.6 & 2.35 (3\%) & 31.98 (44\%) & 31.94 (44\%) & 5.77 (8\%) & 72.05\\
1.0 & 1.46 (3\%) & 21.66 (38\%) & 28.97 (51\%) & 4.36 (8\%) & 56.46\\
2.0 & 0.51 (3\%) & 6.46 (36\%) & 9.87 (54\%) & 1.34 (7\%) & 18.18\\
\hline
\end{tabular}
\caption{Contribution from the different galaxy populations to the bolometric intensity of the extragalactic background light at different redshifts in co-moving frame as defined by Eq.~\ref{bolometric} to the \emph{fiducial} extrapolations (see Sec.~\ref{sec4.1}).}
\label{tab4}
\end{table*}

\subsection{Star formation rate density history}
\label{sec4.3}
Fig.~\ref{fig12} shows the history of the SFR density of the universe computed from our modeling using Eq.~\ref{star}. It is also plotted the prediction using the same equation, from the luminosity densities provided by SGPD10, and a compilation of observational works from \citet{perezgonzalez08} using different estimators, assuming a Salpeter stellar IMF. We are aware that this IMF is not as good description of the observations as other IMFs such as \citet{chabrier03}, but we are concerned here on showing a comparison with the compilation of SFR data, which is given by a Salpeter IMF. The data from $z=3-1.5$ are roughly reproduced. Our results are in agreement within errors with the upper data envelope from $z=1.5-0.7$. We systematically predict a factor $\sim 1.3$ higher SFR than the observational data between $z=0.7-0$. For the \emph{high-starburst} assumption a considerably higher SFR density is estimated. This \emph{high-starburst} case is motivated by the increasing star formation rate density to $z\sim2$ in Fig.~\ref{fig12}, and the increasing specific star formation rate to $z\sim2$ (\citealt{reddy06}; \citealt{daddi07}). But Fig.~\ref{fig12} also indicates that our \emph{high-starburst} is an extreme assumption.

We want to call attention to the large uncertainties on the observational data estimates for the SFR for all redshifts. These uncertainties are especially important for the higher redshifts, mainly because local calibrations are used in the estimations, and also the uncertainties of the corrections due to dust absorption. The same is true for Eq.~\ref{star} which is calibrated using observed local galaxy properties and these might indeed evolve in redshift.

\subsection{Extragalactic background light}
The local EBL ($z=0$) estimated using our method is shown in Fig.~\ref{fig13}. The solid-black line is the EBL calculated by our \emph{fiducial} model\footnote{Intensity files at different redshifts are publicly available at http://side.iaa.es/EBL/} using Eq.~\ref{ebl_spectrum}. Observational data from direct measurements (empty symbols) and from galaxy counts (filled symbols) are plotted. It is usual to consider data from galaxy counts as lower limits. We find a very low background from UV to mid-IR, along the lower limits from galaxy counts. In the UV our model is lower than the \citet{gardner00} data, but we consider these data suspect, due to very poor statistics on their number counts at the faintest magnitudes and the fact that they are systematically higher than the UV data from GALEX (\citealt{xu05}), an experiment with higher sensitivity and better statistics.

In the mid-IR region between 7-15~$\mu$m our results are a factor $\sim1.2$ higher than other models. A lower background than FRV08 is estimated from 15-50~$\mu$m by a factor as large as $\sim 1.5$. Our results are still compatible with the limits from galaxy counts. On the contrary, we predict about the same far-IR light than FRV08 and a factor $\sim 2-3$ larger than GSPD10, higher than the galaxy counts and in very good agreement with most of the direct measurements. The high flux we predicted in the far-IR (in comparison GSPD10) is a characteristic of the SWIRE galaxy SEDs we use, given by the GRASIL code which is used to calculate the far-IR emission, and the relation between the near-IR and the far-IR in the templates.

In the same figure, we also plot upper limits using solid-colour lines from $\gamma$-ray attenuation studies. The cyan and yellow solid lines by \citet{mazin07} were computed for the so-called \emph{realistic} and \emph{extreme} cases, where the authors considered different upper limits for the spectral slopes of VHE emission from  blazars of $E^{-1.5}$ (\citealt{aharonian06}; \citealt{albert08}) and $E^{-2/3}$, respectively. Our calculation is compatible with the upper limits from the extreme case, but marginally disagrees with the realistic case for the largest wavelengths. We will discuss these issues further in Sec.~\ref{attenuation}.

\begin{figure}
\includegraphics[width=\columnwidth]{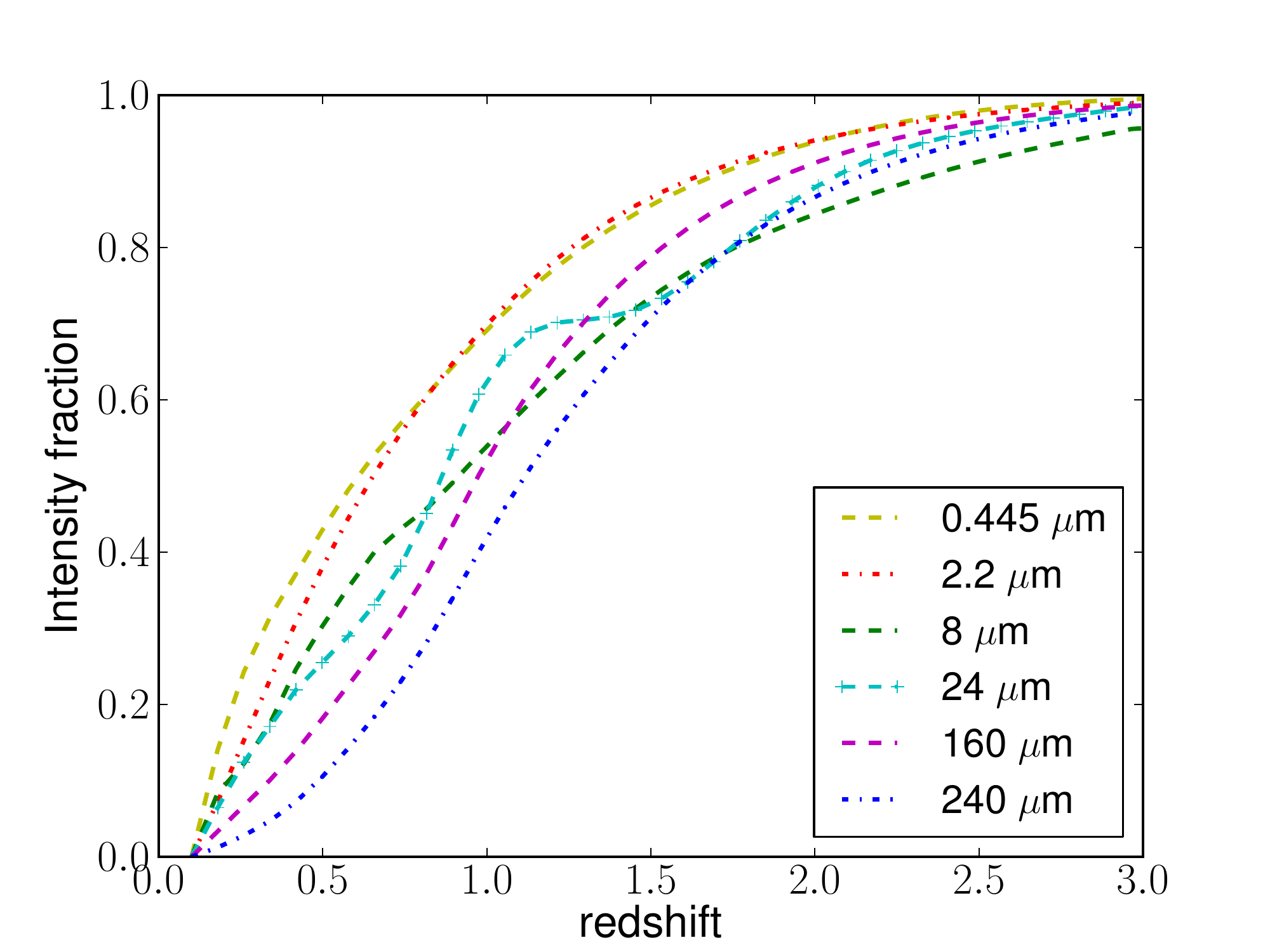}
\caption{Buildup of the extragalactic background light (EBL) at different wavelength normalized to $z=0.1$. For example, according to the \emph{fiducial} model $\sim 70$\% of the local EBL at 2.2~$\mu$m comes from $z<1$, but only $\sim 40$\% of the local EBL at 240~$\mu$m.}
\label{fig15}
\end{figure}

Fig.~\ref{fig13} also shows the uncertainties in our modeling due to the uncertainties on the Schechter parameters of the LF given by C10, the errors in the photometric catalogue, as explained at the beginning of this section, the uncertainties on the $\chi^{2}_{red}$ cut applied, and uncertainties due to the extrapolations for the galaxy-SED types for $z>1$. All the possibilities are calculated and the extreme cases are plotted. The uncertainties from the UV up to the mid-IR are dominated by the errors in the photometry and the cuts. The directions from both effects are different: the uncertainties from the photometry are below the \emph{fiducial} model, the uncertainties from the cuts are above it. In the far-IR the uncertainties in the extrapolations to $z>1$ dominate with a factor $\sim 1.5$. These effects will be thoroughly discussed in Sec.~\ref{sec6.1}.

The evolution of the EBL is important to account for the history of the galaxy emission and the processes involved, as well as to properly calculate the attenuation for VHE $\gamma$-rays for the high-redshift universe. We show, in Fig.~\ref{fig14}, the co-moving intensity level of the EBL for different redshifts, the contribution to the EBL at those redshifts from the four main SED groups to our \emph{fiducial} extrapolation, and the predictions for the EBL by other models. In Table~\ref{tab4} we quantify this evolution, where the bolometric intensity is defined according to Eq.~\ref{bolometric}, \ie
\begin{equation}
\label{bolometric}
I_{bol}=\int\nu I_{\nu} d\ln{\nu}\hspace{3cm}
\end{equation}

We should note that the starburst population contributes 54\% to the co-moving bolometric EBL at $z=2$, but only 30\% for the local universe. We note as well that the far-IR peak in the SED is higher relative to the near-IR peak at these redshifts; this is due to the fact that a large fraction of the energy radiated from starburst systems is at far-IR wavelengths. We also note that the total bolometric intensity peaks at $z\sim0.6-0.2$, because the far-IR peaks at higher energetic wavelengths there as seen in Fig.~\ref{fig14}.

\begin{figure}
\includegraphics[width=\columnwidth]{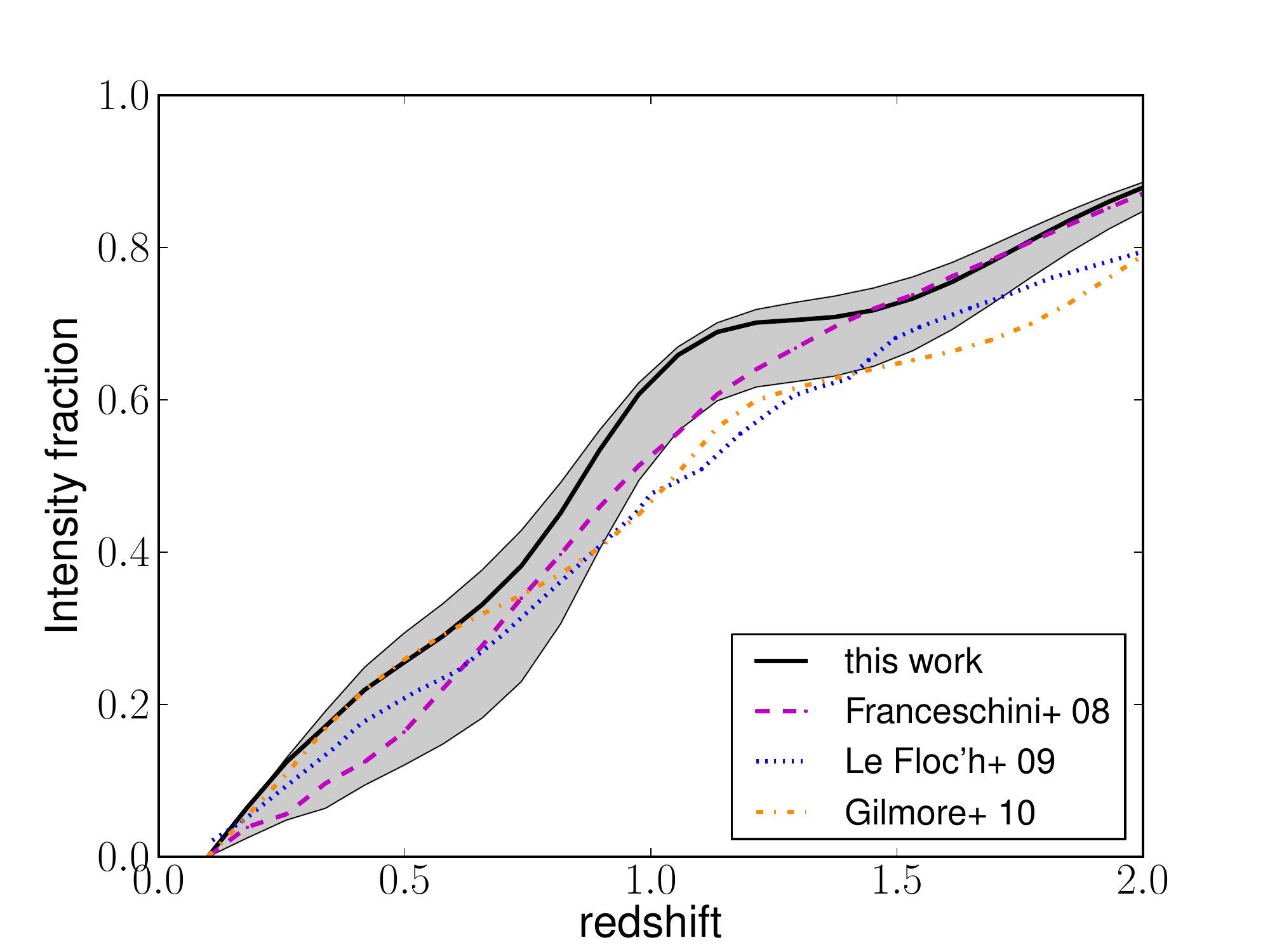}
\caption{Buildup of the extragalactic background light (EBL) at 24~$\mu$m obtained from different phenomenological models, normalized to $z=0.1$, compared with the Spitzer/MIPS data by \citet{lefloch09}. For example, according to our \emph{fiducial} extrapolation (see Sec.~\ref{sec4.1}), about 75\% of the local EBL at 24~$\mu$m was already in place at $z\sim$1.5. Uncertainties in the modeling are shown with a shadow region (see Fig.~\ref{fig13}). The curve from Franceschini et al. (2008) has been calculated by us from their published EBL densities.}
\label{fig16}
\end{figure}

Another important observable is the buildup of the local intensities for different wavelengths. This is the fraction of the local EBL at a given wavelength that was already in place at a given redshift. This is shown in Fig.~\ref{fig15} for several wavelengths. As an example, we see that $\sim$70\% of the local EBL at $\lambda=0.445~\mu$m and 2.2~$\mu$m comes from $z<1$, 50\% of the EBL below $\sim$180~$\mu$m was already in place at $z=1$, but it is only $\sim 40$\% at 240~$\mu$m. It is significant that the EBL at shorter wavelengths mostly come from sources at much lower redshifts than the larger ones (see \citealt{lagache05}).

Fig.~\ref{fig16} shows a comparison between the EBL buildup for our model, FRV08, GSPD10, and the observational work by \citet{lefloch09} based on data from MIPS at 24~$\mu$m up to $z\sim$~1.5 in the COSMOS field. The main contribution to the EBL at 24~$\mu$m comes from star-forming and starburst-type galaxies. This region of the SEDs is highly dependent on the non-smooth PAH features. We observe a general agreement, but reaching a factor 40\% difference at $z\sim1.2$ for the \emph{fiducial} extrapolation. The uncertainties here are large (see Sec.~\ref{sec6.1}).

\section{Gamma-ray attenuation}
\label{attenuation}

\subsection{$\gamma$-ray attenuation from this EBL model: theoretical background}
The EBL has important implications for the interpretation of data taken using recent VHE experiments (the Fermi satellite, \citealt{gehrels99}; and IACTs, such as MAGIC, VERITAS and HESS; \citealt{lorenz04}; \citealt{weekes02}; \citealt{hinton04}, respectively), due to the photon-photon pair production between $\gamma$-ray photons traveling across cosmological distances and EBL photons (see \citealt{nikishov62}; \citealt{gould66}).

Blazars are an important source of extragalactic $\gamma$-ray emission and have become a relevant tool for indirectly measuring the EBL. These objects are believed to be an extreme category of AGNs. Their emission, which occurs at all wavelengths of the electromagnetic spectrum, comes from supermassive black holes (with masses $\geq$~10$^{7}$~M$_{\sun}$) swallowing matter accreted from their surroundings. In general, AGNs are characterized by a beamed emission perpendicular to the accretion disc known as jets, which are pointing toward us in the case of blazars.

The current theoretical models for the emission by this class of objects are of two kinds: leptonic or hadronic. Both models predict a spectrum with two peaks, the first one localized from radio to $X$-rays due to synchrotron radiation from relativistic electrons (leptonic model), or protons (hadronic model). However, the second peak has a different nature.  While in the leptonic model it is due to inverse Compton (IC) scattering of the same population of electrons that produce the synchrotron peak (\citealt{boettcher07}), in the hadronic model, nuclear photo-disintegration is advocated to explain the second peak (\citealt{sikora09}). Both models face serious difficulties in explaining intrinsic (\ie EBL-corrected) VHE power law indices harder than 1.5, and fail to explain slopes harder than 2/3. The intrinsic spectrum is the spectrum that we would observe if there were no effect from the EBL.

\begin{figure}
\includegraphics[height=10cm,width=\columnwidth]{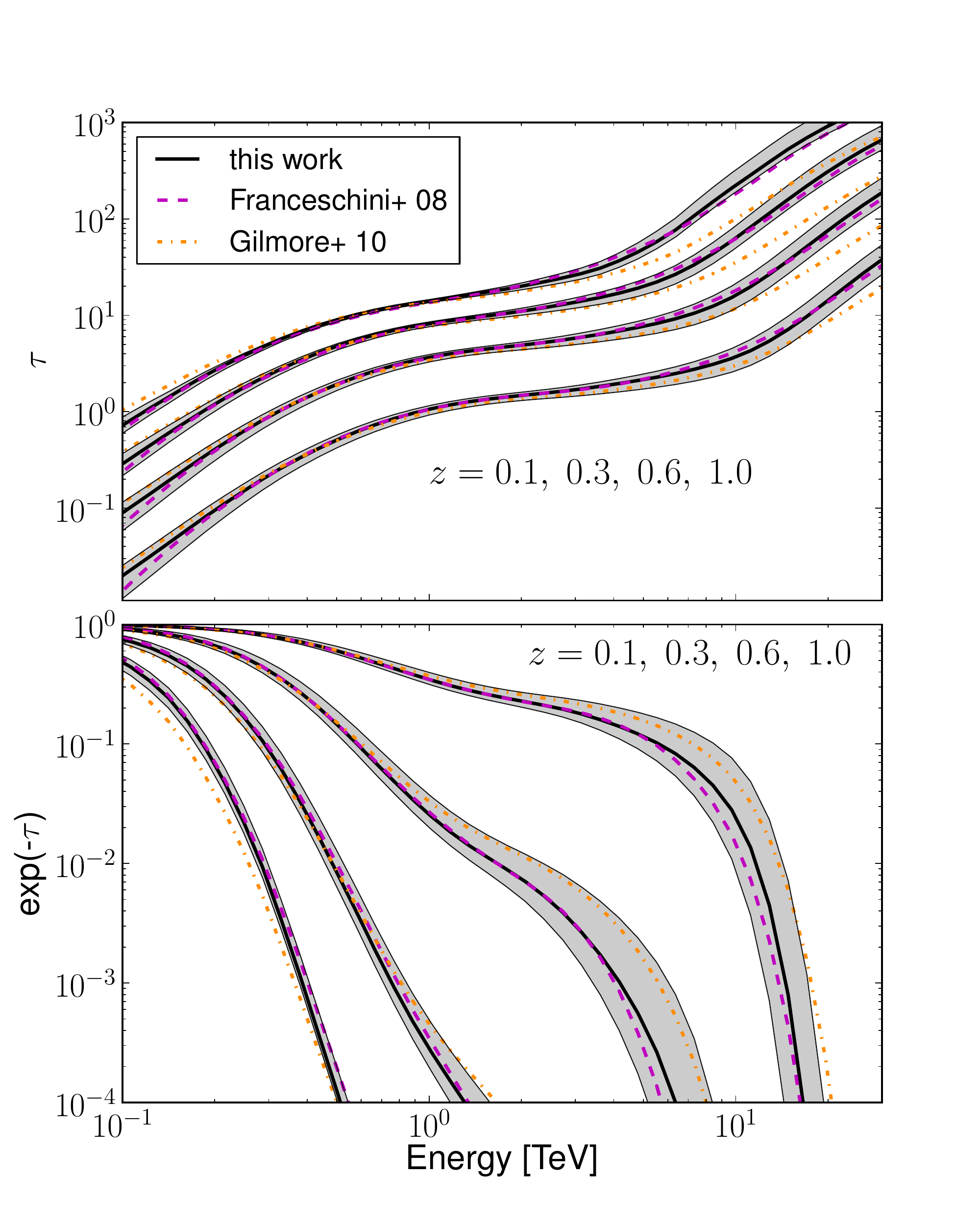}
\caption{\emph{Upper panel}: Optical depth versus observed energy of $\gamma$-ray photons for sources at different redshifts (from bottom to top $z=0.1$, 0.3, 0.6 and 1), due to the extragalactic background light computed for our model in solid-black line, for Franceschini et al. (2008) in dashed-magenta line, and for \citet{gilmore10b} in orange-dot-dashed line. \emph{Lower panel}: Flux attenuation versus observed energy of $\gamma$-ray photons for fictitious sources at different redshifts (from right to left $z=0.1$, 0.3, 0.6 and 1). We have calculated attenuation for the Franceschini et al. (2008) and \citet{gilmore10b} models using the EBL data provided by the authors. The EBL uncertainties in Fig.~\ref{fig13} are propagated to the optical depth and flux attenuation. They are shown here with a shadow region.}
\label{fig17}
\end{figure}

The EGRET satellite observed AGNs in the local universe (hence not very attenuated), claiming that all of them have spectral indices $E^{-\Gamma_{int}}$ with $\Gamma_{int}\ge 1.5$ in the high energy (30~MeV-30~GeV) regime (\citealt{hartman99}). This result has been confirmed by the Fermi Collaboration (within uncertainties), which has published a catalogue of AGNs detected by the Fermi Large Area Telescope (LAT) all-sky survey during its first year in operation (\citealt{abdo10}). From this observational fact, and the theoretical issues above, it is usually conservatively consider that no AGN could have an intrinsic VHE spectrum fitted by a power law with an index harder than 1.5. Some authors such as \citet{katarzynski06}, \citet{stecker07}, \citet{bottcher08} and \citet{aharonian08} provide some mechanisms within standard physics to reach slopes harder than 1.5, but never harder than $\Gamma_{int}=~2/3$.

The EBL may be constrained using VHE observations of extragalactic sources if their intrinsic emitted spectra are known. As mentioned in Section \ref{intro}, $\gamma$-ray photons coming from cosmological distances are attenuated by photon-photon pair production by EBL photons. The cross section of this reaction depends of the product shown in the left side of Eq.~\ref{cross},
\begin{equation}
\label{cross}
\sqrt{2\varepsilon E(1-\cos{\theta})}\ge 2m_{e}c^{2}
\end{equation}

\begin{equation}
\label{threshold}
\varepsilon_{th}\equiv \frac{2m_{e}c^{2}}{E(1-\cos{\theta})}
\end{equation}
\noindent where, in the rest-frame at redshift $z$, $E$ is the energy of the $\gamma$ photon, $\varepsilon$ is the energy of the EBL photon, and $\theta$ is the angle of the interaction, which defines an energy threshold $\varepsilon_{th}$ for the EBL-photon energy given in Eq.~\ref{threshold} with $m_{e}$ the electron mass.

The cross section peaks at about twice $\varepsilon_{th}$, which produces a peak in the interaction at $\lambda$~[$\mu$m]~=~1.24$E$~[TeV]. From this property, a $\gamma$-ray with energy 1~TeV interacts mainly with a photon of the EBL with wavelength $\sim$~1 $\mu$m. The details may be found for example in \citet{madau96}.

For a given observed spectrum of a source at redshift $z$ we can find the intrinsic spectrum by assuming a particular EBL model and multiplying by the attenuation factor to \emph{de-absorb} the spectrum using Eq.~\ref{correct}, \ie

\begin{equation}
\label{correct}
\frac{dF}{dE}\Big{|}_{int}=\frac{dF}{dE}\Big{|}_{obs}\exp{[\tau(E,z)]}
\end{equation}
where the subscript $obs$ means observed, $int$ is intrinsic, and
$\tau(E,z)$ is the optical depth dependent on the observed energy $E$ of the $\gamma$
photon for a given EBL photon density and redshift,

\begin{equation}
\label{attenu}
\tau(E,z)=\int_{0}^{z} \Big(\frac{dl'}{dz'}\Big) dz' \int_{0}^{2}d\mu \frac{\mu}{2}\int_{\varepsilon_{th}}^{\infty} d\varepsilon{'}\ \sigma_{\gamma\gamma}(\beta{'})n(\varepsilon{'},z')
\end{equation}

\begin{equation}
\label{beta}
\beta^{'}=\frac{2m_{e}^{2}c^{4}}{E\varepsilon\mu(1+z)^{2}}
\end{equation}
with $dl'/dz'=c|dt'/dz'|$ given by Eq.~\ref{peebles}, $\mu=1-\cos\theta$, $\sigma_{\gamma\gamma}$ the photon-photon pair production cross section, $\beta{'}$ is given by Eq.~\ref{beta} and $n$ is the proper number density per unit energy of EBL photons\footnote{Attenuation files are publicly available at http://side.iaa.es/EBL/}. We show in Fig.~\ref{fig17} the optical depth and attenuation for sources at $z=0.1$, 0.3, 0.6 and 1.

Since the EBL produces an attenuation of the VHE spectra, a mere detection of VHE photons (using some constraint on the intrinsic blazar power spectrum) places an upper limit on the EBL density. Some upper limits have been derived by different authors, fitting EBL models to the density level where the condition $\Gamma_{int}=1.5$ is satisfied, building \emph{ad-hoc} EBL models. We plotted those limits in Fig.~\ref{fig13} (\citealt{aharonian06}; \citealt{mazin07}; \citealt{albert08}). Each of those upper limits comes from the study of different blazars with a different measured energy spectrum. Due to the peak of the interaction previously mentioned, each of the studies constrains different ranges on the EBL. \citet{aharonian06} used the VHE spectrum of the blazar 1ES~1101-232 at $z=0.186$ observed from 0.2-3~TeV, scaling the model by \citet{primack01} multiplying the total EBL intensity by a constant to satisfy the $\Gamma_{int}=1.5$ condition. \citet{albert08} used the spectrum of 3C~279 at $z=0.536$ observed from 0.08-0.5~TeV, scaling a slightly modified model by Kneiske et al.~(2002). \citet{mazin07} used a compilation of blazars at different redshifts and observed at different energies, and splines from a grid as EBL densities. They make two different assumptions about the maximum $\Gamma_{int}$ leading to two different upper limits (called by the authors \emph{realistic} and \emph{extreme}).

We see in Fig.~\ref{fig13} that the \emph{fiducial} EBL model (hereafter all the results in this section are discussed for this, unless otherwise stated) is below the upper limits at all wavelengths, except at the largest wavelengths, where slightly exceeds the limits from the \emph{realistic} case by \citet{mazin07}. This fact is discussed in Sec.~\ref{mrk501} and it is explained why we do not consider this a major problem. Another limit not plotted comes from the blazar 1ES~0229+200 at $z=0.140$ (\citealt{aharonian07c}). Its study set a lower limit in the slope of the local EBL spectrum between 2-10~$\mu$m, $\alpha$~$\geq$~1.10~$\pm$~0.25, to satisfy the limit on AGN's spectra $\Gamma_{int}$~$\geq$~1.5. We remark that they set the limit only on the slope, not on the intensity level. We have fitted our model in that wavelength range, to a power law $\propto\lambda^{-\alpha}$ getting $\alpha=1.19$~$\pm$~0.07. Our model is thus compatible with this constraint.

\begin{figure*}
\includegraphics[width=18cm]{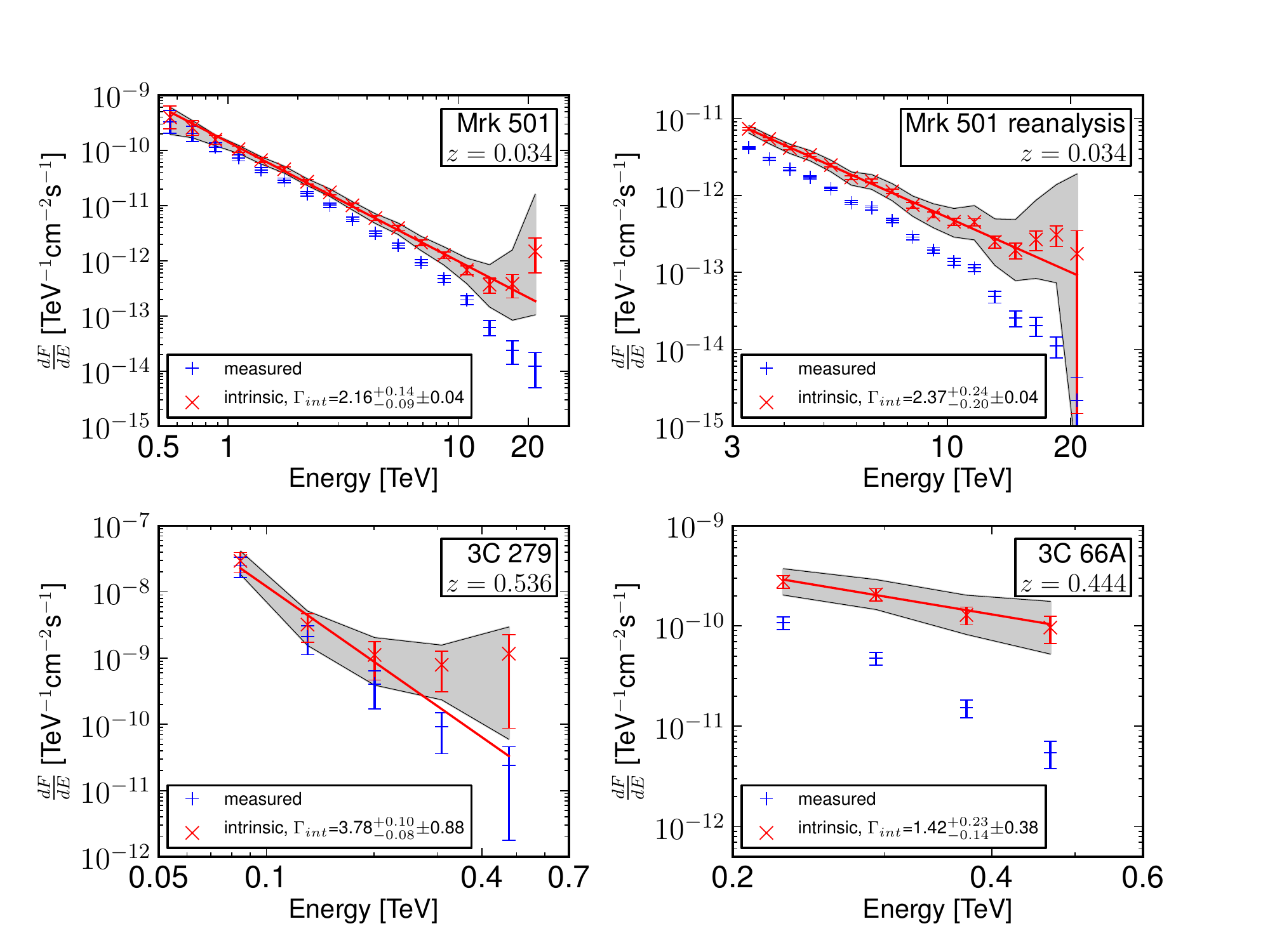}
\caption{Very high energy spectra measured (blue) and EBL-corrected from the attenuation calculated with our EBL model (using the \emph{fiducial} extrapolation for the galaxy SED-type fractions at $z>1$, in red) of three extreme blazars: Mrk~501 observed in very high state up to energies larger than 20~TeV (upper left panel, \citealt{aharonian99}) and a reanalysis of the same data (upper right panel, \citealt{aharonian01}), 3C~279 a flat-spectrum radio quasar with the highest redshift ($z=0.536$) ever detected for a VHE $\gamma$-ray source (lower left panel, \citealt{albert08}) and 3C~66A a BL Lac with probably (because its redshift, $z=0.444$ is not very secure) the highest redshift ever detected for an object of this class (lower right panel, \citealt{veritas09}). Uncertainties from the EBL modeling as well as statistical and systematic errors are shown with a shadow region. The straight red line is the best-fitting power law for every blazar with index $\Gamma_{int}$. The first uncertainties in the index are due to the EBL modeling as shown in Fig.~\ref{fig17}, and the second uncertainties are statistical plus systematic errors in all blazars, except 3C~66A where only statistical errors are shown.}
\label{fig18}
\end{figure*}

It is also possible to set upper limits on the unknown redshift of blazars assuming an EBL model and finding the redshift by which the EBL-corrected spectrum satisfies $\Gamma_{int}=1.5$ (\citealt{prandini10}; \citealt{yang10}). We apply that method to the PG~1553+133 spectrum observed by MAGIC (\citealt{albert07}), assume an EBL-corrected spectrum given by a power law and find an upper limit at $z\leq 0.85\pm 0.07$ in agreement with the lower limit ($z\geq 0.4$) found by \citet{danforth10} using absorption features in the Ly-$\alpha$ forest of the blazar.

As shown in Fig.~\ref{fig17}, our EBL model implies about the same attenuation as other recent models we compare to over all the energy range observed by the current generation of IACTs. Larger transparency than the observationally-based model by FRV08 is found (roughly a factor $\sim2$ in flux, but still within the uncertainties) for $\gamma$-ray photons with energies between $\sim 6-15$~TeV for $z\sim 0.1$, but a factor $\sim 2$ in flux less transparent than the GSPD10 theoretical approach around $\sim 10$~TeV. For the large-redshift case, our model predicts about the same attenuation as FRV08, but a factor $\sim 1.5$ more transparency than GSPD10 for sub-TeV energies. Note that a small difference in the optical depth has large effects on the spectra due to the exponential in Eq.~\ref{correct}, \eg a factor 1.5 in optical depth leads to a factor $\sim 5$ in attenuation.

\subsection{Application to extreme known blazars}
We now proceed to test whether the observed spectra of the three most constraining AGNs known in the VHE range due to their hard spectrum, or to their large redshift, satisfy the condition that the intrinsic spectrum corrected by the attenuation derived with our model has $\Gamma_{int}$~$\geq$~1.5. We consider the blazars: Mrk~501 at $z=0.034$ detected by the HEGRA system of Cherenkov telescopes in 1997 (\citealt{aharonian99}, with a reanalysis by \citealt{aharonian01}), FSRQ 3C~279 at $z=0.536$ observed by MAGIC (\citealt{albert08}), and the blazar 3C~66A observed by VERITAS at $z=0.444$ (\citealt{veritas09}), all of them seen in a flaring state. All these blazars are plotted in Fig.~\ref{fig18} showing in the legends that the condition $\Gamma_{int}\ge1.5$ is satisfied.

\subsubsection{Mrk~501}
\label{mrk501}
The highest energy bins in this measurement, where it is observed a significant deviation from a power law (see Fig.~\ref{fig18} upper-left panel), are affected by the far-IR EBL at $\lambda>60~\mu$m.  This is the region of the EBL spectrum where it was found a disagreement with the \emph{realistic} (but not \emph{extreme}) upper limits of \citet{mazin07}. The problem comes from the very low statistics and high systematic uncertainties at such high energies (\citealt{aharonian99}). A later reanalysis of the same observation done in \citet{aharonian01} accounts for larger systematic uncertainties as shown in Fig.~\ref{fig18} upper right panel.

This exponential behaviour for the highest energy bin was already observed from the first EBL models (\eg \citealt{malkan98}; \citealt{primack99}; Kneiske et al.~2002), whose EBL levels were higher than the more recent ones. This fact was discussed thoroughly in \citet{dwek05}, and even some exotic explanations such as Lorentz invariance violation (\citealt{stecker01}) were proposed. More recent EBL models with a more transparent universe (such as our model, FRV08 and GSPD10) relax such predictions. The solutions to exponential spectra and photon pileup could involve widespread problems with the photon statistics and systematic uncertainties in the observations (as suggests the results from the later reanalysis), or new mechanisms extending the normal SSC model, using external regions close to the $\gamma$-ray source with target photons. The EBL uncertainties in the far-IR leading to the attenuation uncertainties at these high energies as shown in Fig.~\ref{fig18}, might contribute to the solution as well.

Another observed flare with better statistics with the current generation of IACTs up to such high energies as $\sim 20$~TeV would be very helpful in constraining these possibilities.

\subsubsection{3C~279}
Fig.~\ref{fig18} shows in the lower-left panel the EBL-corrected VHE spectrum for this source. An external photon field providing target photons for IC (such as that provided by a broad-line emission region) might be necessary to explain the flat behavior at the largest energy bins, as discussed in \citet{albert08}. Instrumental systematic uncertainties might explain this behavior as well. We note here that our model is already matching the lower limits from galaxy counts at the wavelengths where $\gamma$-ray attenuation with the observed energies occurs, and a much lower EBL density than the one calculated in this work does not seem realistic. The attenuation uncertainties from the EBL modeling are too low at this redshift and these energies to explain that spectral behaviour.

\subsubsection{3C~66A}
Fig.~\ref{fig18} shows in the lower-right panel the EBL-corrected VHE spectrum for this source, whose EBL-corrected slope is well within the 1$\sigma$ limit of the $\Gamma_{int}\ge 1.5$ according to the calculated intrinsic index. It is important to note that the redshift considered for this object is calculated using just one emission line and is thus not very secure (see discussion in \citealt{bramel05}), and its attenuation might be indeed overestimated if the redshift is lower than assumed.

\subsection{Conclusions on the limits from blazars}
It is concluded from the study of these extreme blazars that our EBL is generally compatible with the hardness of the EBL-corrected slopes expected from theoretical arguments. However, it is clear that a simple SSC model cannot explain any flatness at the highest energies of the EBL-corrected spectra of either Mrk~501 or 3C~279, which suggests that some extension to the model may be necessary such as an external photon region, a better understanding of the IACT systematic uncertainties, or even a revision of the propagation mechanisms mainly through the intergalactic medium (see \citealt{sanchezconde09}).

\subsection{Propagation of the EBL uncertainties to the $\gamma$-ray attenuation}
As shown in Fig.~\ref{fig17} the uncertainties in the attenuation are dependent on the observed $\gamma$-ray energy as well as the redshift: the higher the energy or the redshift, the higher the uncertainties in the attenuation. The attenuation uncertainties shown were calculated from the uncertainties in our EBL modeling, which were shown in Fig.~\ref{fig13}, explained in Sec.~\ref{results} and will be thoroughly discussed in Sec.~\ref{sec6.1}. For sub-TeV energies up to around 1~TeV the uncertainties in the flux attenuation are never higher than a factor $\sim 2$ and generally lower. The uncertainties in the EBL-corrected spectra in this case are dominated by other effects (see indices in Fig.~\ref{fig18}). For energies larger than 10~TeV the uncertainties are around a factor of several. The uncertainties in the EBL-corrected spectra up to such high energies due to the EBL modeling are considerable. These high uncertainties are derived from the EBL in the far-IR region due to the very fast increment of the EBL photon density ($n$, see Eq.~\ref{attenu}) with longer wavelengths. Observations of sources at low redshift but energies larger than $\sim 10$~TeV will set constraints on these uncertainties.

\section{Discussions and comparison with semi-analytic models}
\label{discussion}
\subsection{Discussion on EBL uncertainties}
\label{sec6.1}
As explained in Sec.~\ref{method}, we adopt a lower limit to the photometric errors higher than those in the AEGIS catalogue. Different lower limits are set from 1-10\% of the photometric measurements. That is, if the error in any band is lower than our limit then we set it to the limit. The results are sensitive to the limit choice in $R$ and $I$ (where the errors in the catalogue are the lowest), but not for the other bands. The galaxy SED-type fractions change for lower limits 1-6\%, but there is little change if the level is set higher than 6\%. The change is mostly for the quiescent and star-forming galaxy fractions. If we use the errors in the catalogue without any change, we find 10\% more quiescent galaxies at $z=0.3$ than for a lower limit of 6\%, which decreases to $\sim 3\%$ more quiescent galaxies at $z=0.9$ than for a lower limit of 6\%, as shown in Fig.~\ref{fig8}. The change is mostly in Ell2-type galaxies, according to the SWIRE classification. We have investigated those quiescent galaxies that change their best fit to star-forming galaxies upon raising the lower limit on the errors, and find that they are often fitted much better by a star-forming SED. In many cases they even have detection in MIPS~24, clearly indicating ongoing star formation. On the other hand, based on the comparison of our photometric measurements to those of other catalogues we estimate that any error in the photometry lower than $\sim 5\%$ should not be considered very reliable. For those reasons we set the lower limit at 6\% for model. The uncertainties due to this are below the \emph{fiducial} model in Fig.~\ref{fig13}, and for the reasons stated an EBL in this region should not be considered very likely (and therefore, neither is their derived attenuation in Fig.~\ref{fig17}).

Another source of uncertainty accounted for in Fig.~\ref{fig8} and Fig.~\ref{fig13} is the $\chi^{2}_{red}$ cut that separates good and bad SED fits. The main change occurs for AGNs, where for a relaxation in the cut (from $\chi^{2}_{red}=10$ to 20), the fraction can increase by as much as 10\% at $z=0.9$, and by a smaller fraction at $z=0.3$. These changes affect the EBL in the following way: higher AGN fractions increase the UV as well as the mid-IR, while higher quiescent fractions decrease the flux at those wavelengths. This effect affects the uncertainties above the \emph{fiducial} model and an EBL intensity in this shadow region is considered more reliable than the region below the \emph{fiducial} model (the same for their derived attenuation in Fig.~\ref{fig17}). While $\chi^{2}_{red}$ cuts do not have an appreciable effect on the far-IR flux, there is a substantial change arising from the choice of extrapolation in SED-types above $z\sim1$, as we find in the \emph{high-starburst} assumption. Fig.~\ref{fig13} also accounts for the uncertainties in the Schechter parameters of the LF given by C10 but these are small.

Two major potential problems for our modeling might be a colour-dependent selection effect and the extrapolation of the galaxy SED-type fractions for $z>1$. It was already shown in Sec.~\ref{sec3.2} that the colour-selection effect is rather small. From the fact that most of the light in the EBL comes from the knee of the LF around $L_{\star}$, where our sample does not suffer any color-dependent selection effect, we do not consider this to be a significant problem for our EBL calculation. Our estimated galaxy SED-type fractions appear to be consistent with works by others as well. For example, our results agree with \citet{blanton06} and \citet{faber07}, who find roughly no evolution for late-type (blue) galaxies from $z\sim1-0$ within a 10\% range, and an increment of the early-type (red) population in the same redshift range by at least a factor 2. We also highlight that the galaxy SED-type fractions that we calculated for the local universe smoothly link with our independently-derived results at $z\sim0.3$.

Regarding the galaxy SED-type fraction extrapolations, we have considered two rather different approaches which basically lead to the same evolving luminosity densities and EBL for the optical/near-IR range where $\gamma$-ray attenuation occurs, as shown in Fig.~\ref{fig11} and Fig.~\ref{fig14}, respectively. This fact is due to the shape of the stellar emission, because the contribution to the optical/near-IR peak is very similar for quiescent, star-forming and starburst galaxies for a given $M_{K}$. We recall here that the normalization to our model is fixed by the rest-frame $K$-band LF by C10. The only difference between our extrapolations is at the far-IR peak, where our results are considered to be less robust for this reason, as well as for the reasons stated below regarding the SWIRE templates and the lack of photometric data. Deeper observations by future galaxy surveys will help in characterizing the galaxy SED-fractions up to higher redshifts.

It was also checked how the fractions change if the detection limit is relaxed from $5\sigma$ to $3\sigma$ for the bands where there are observations, but no detections. Many more quiescent galaxies than in the $5\sigma$ case were found, even a factor larger than 2, due to the low detection limit on the MIPS~24 instrument, but generally they are not good fits.

In our work we have not differentiated between the spectroscopic and secure photometric redshifts. This is an approximation, and it is necessary to check that this is consistent with our results. We find that the galaxy SED-type fractions derived from both sub-catalogues are clearly compatible and the trends are the same.

Galaxies fitted to a starburst SED may instead be very late-type star-forming galaxies (or viceversa), because both SED templates are rather similar in the regions where we have data. This may be called mis-typing and its effect is expected to be larger for faint galaxies, because the major fraction of faint galaxies are star-forming or starburst and not massive quiescent or AGN galaxies. Such small galaxies are probably rather metal poor and thus lacking dust. Hence, their SEDs are probably more like star-forming galaxies rather than starbursts (in agreement with our results in Fig.~\ref{fig9}). It is a source of uncertainty in the mid and far-IR (underestimating or overestimating light) and might explain the excess found in Fig.~\ref{fig11} and \ref{fig12} compared with the data. Far-IR data would help in resolving this issue, but the number of galaxies with detection in MIPS~70 is rather low to make statistical estimations. Herschel data will be very useful thanks to its good spatial resolution and deep photometry in the far-IR.

Another source of uncertainties in our model that quantitatively we have not accounted for arises from the use of local SED templates to fit galaxies at $z>0.3$. This comes from the fact that the SWIRE templates are based on observations of local galaxies, and we expect that they become worse fits when the redshift is increasing. This problem will be addressed by new data from WFC3 on Hubble Space Telescope and the next generation of ground and space optical/near-IR telescopes such as the James Webb Space Telescope.

The lowest EBL flux in Fig.~\ref{fig13} is given by the case with the highest number of quiescent galaxies and lowest number of AGN galaxies, which corresponds to the case of using the low errors in the catalogue and our $\chi^{2}_{red}$ cuts. The highest EBL flux occurs with fewer quiescent galaxies and the highest fraction of AGN galaxies. This is the case with the 6\% lower limit for the errors in the photometry and without a $\chi^{2}_{red}$ cut. Using the lower limits from galaxy counts in the UV and in the optical we may rule out at $>2\sigma$ the mixing of galaxy SED-type fractions predicting the highest fractions of quiescent galaxies and fewest AGNs in Fig.~\ref{fig8}. We do not consider the VHE observations to exclude the models with higher far-IR, because the discrepancy is for wavelengths longer than 60~$\mu$m where those limits may not be reliable for the reasons stated in Sec.~\ref{attenuation}. Further VHE observations might indeed constrain our galaxy SED-type fractions.

Thus to recap, the EBL uncertainties from the UV up to the mid-IR are low enough to recover the spectra of $\gamma$-ray sources with energies lower than $\sim 10$~TeV, but the EBL uncertainties have to be reduced in the far-IR (for neglecting uncertainties due to the EBL modeling) to correct higher energy sources. Additional photometry is needed there to clearly distinguish between star-forming and starburst galaxies, therefore to reduce the mis-typing, as well as a better understanding on the far-IR region of the galaxy SEDs at $z>0.3$. Characterizing the galaxy SED-type fractions at $z>1$ will reduce these uncertainties in the far-IR as well.

\subsection{Discussion of the results}
The local luminosity density from galaxies is observationally well constrained over all wavelengths from 0.1-1000~$\mu$m. As shown in Fig.~\ref{fig10}, the prediction of the local luminosity densities is in very good agreement with observational results.

Fig.~\ref{fig11} showed the evolving rest-frame UV luminosity density as well. These results agree well with the observational data by \citet{gabasch06} and \citet{dahlen07} within uncertainties, but they are a factor $\sim 1.6$ below the data for $z=1.7$ and 2.2. It was also compared in Fig.~\ref{fig11} the evolution of the rest-frame luminosity in the $K$-band from our calculations to independent observational works. Some disagreement was found that in the case of \citet{barro09} might be explained by the fact that they do not correct their sample for incompleteness and only consider the brightest sources, unlike the LF by C10. Therefore their results should be considered as lower limits. The direct comparison in Fig.~\ref{fig11} with the rest-frame luminosity density in $B$-band showed that our luminosity in that band is not in contradiction with other independent works. We are doing really well reproducing the data from \citet{norberg02}, \citet{gabasch04} and \citet{ilbert05}. We might be indeed overestimating the light in this band 15-20\% for $z<1$ according to the data by \citet{faber07} and a factor $\sim2$ for $z>1$ according to the data by \citet{dahlen05}, but this latter does not significantly affect our results for the local EBL because as we already showed in Fig.~\ref{fig15}, most of the optical/near-IR light comes from $z<1$.  The comparison with the bolometric IR luminosity density with the observational works by FRV08 and \citet{rodighiero10} is very good, even though we are a factor $\sim 2$ higher around $z\sim1$.

A good agreement was found with the upper envelope of the data cloud on the calculations for the SFR history from $z\sim1.5$ down to the local universe using Eq.~\ref{star} (see Fig.~\ref{fig12}), using our \emph{fiducial} extrapolation. According to \citet{magnelli09} at around $z\sim1$ the main contributor to the star formation is the obscured IR contribution, instead of the UV. We may be overpredicting some of this obscured IR light around $z\sim1$ due to the lack of far-IR photometric data in our galaxy catalogue that allow a clear classification between late star-forming and starburst galaxies as discussed in Sec.~\ref{sec6.1}. With the \emph{high-starburst} extrapolation of the galaxy-SED-type fractions was checked that increasing the starburst-like population a factor $\sim3$ from $z\sim1-2$, we may get a flatter SFR density history up to $z\sim2$, but even higher than the observational data. This \emph{high-starburst} assumption does not change our general picture of the local EBL, but increases the far-IR peak a factor $\sim 1.5$ (as it was considered in Fig.~\ref{fig13} and was shown explicitly in Fig.~\ref{fig14} for some other redshifts).

The EBL calculated in this work is matching the data from galaxy counts from the UV up to the mid-IR (see Fig.~\ref{fig13}), except the data found in \citealt{levenson08}. Higher intensities than the data from galaxy counts were calculated in the far-IR but in agreement with direct detections. The EBL evolution shown in Fig.~\ref{fig14} is in good agreement with FRV08 up to $z=1$. At higher redshift our results are different in the UV and optical/near-IR. This may be due to the fact that FRV08 extrapolate the galaxy evolution, while in our model this evolution is entirely based on the observed LF by C10 up to $z=4$. See Sec.~\ref{SAMs} for a comparison with the results by GSPD10.

There are some works in the literature where the contribution from AGN galaxies to the EBL is studied. According to recent works that focus in the mid-IR (\eg \citealt{silva04}; \citealt{matute06}) this contribution should not be larger than 10-20\%. This is in agreement with our results: we find that the AGN-galaxy contribution to the bolometric EBL is 6\% for the \emph{fiducial} extrapolation (Table~\ref{tab4}) and 13\% for the case with the largest AGN fraction in Fig.~\ref{fig8}. For the wavelength range between 1-20~$\mu$m the AGN contribution from our model is also between 8-16\%. We estimate that this contribution to the co-moving bolometric EBL slightly increases with redshift.

The EBL buildup was studied in Fig.~\ref{fig15} and \ref{fig16}. It was found that most of the local UV/optical/near-IR EBL was built up at $z<1$, while the far-IR EBL was mostly built up at $z>1$. This result for the far-IR light agrees well with the observational work by \citet{devlin09}, but disagrees with \citet{chary10}. In any case, our uncertainties in the far-IR are very large. Differences up to 40\% were found in the buildup of the local EBL at 24~$\mu$m. These differences are due to the fact that a very small change in the mid-IR region leads to a very strong difference in this buildup plot, and to the mid-IR peak that we get at larger redshifts (see Fig.~\ref{fig14}) due to the shape of the galaxy SEDs. We point out here that the EBL buildup is on how the light is being built up, and not about the absolute intensity value.

We already discussed our results on $\gamma$-ray attenuation in Sec.~\ref{attenuation}

\subsection{Comparison with SAMs}
\label{SAMs}
In this section we compare our EBL estimation against the EBL model described in SGPD10 and GSPD10, which is based on SAMs of galaxy formation. The comparison for $\gamma$-ray attenuation has been thoroughly discussed in Sec.~\ref{attenuation}. We notice that slightly different cosmological parameters were used for our model and that by SGPD10. The latter uses the latest values from WMAP5, which slightly affect all the results in the local universe as well as their evolution.

We already saw in Fig.~\ref{fig11} the comparison between our observational luminosity densities and the theoretical prediction by SGPD10 for the co-moving luminosity density versus redshift in the UV, in the near-IR ($K$-band), in the optical ($B$-band) and for the bolometric IR luminosity. We note that our $K$-band luminosity density evolution is given exclusively by the C10 LF, because at that band our choice of galaxy SED fractions does not affect our results. This quantity is above the prediction by SAMs by a factor around 20\% from $z\sim2$ down to the present universe. The UV from this SAM is above our results for all redshift, except at $z\sim 1$. At $z\sim 2$ is a factor 4 higher. For the $B$-band luminosity density the agreement is excellent from $z\sim1$ down to the local universe. For $z>1$ SAMs predict a factor of several more light than our observationally-based approach. We may see the consequences of this for the EBL evolution in Fig.~\ref{fig14} for high redshifts where the excess of light has not been diluted by the expansion of the universe. For the bolometric IR luminosity SAMs seem to systematically predict at least a factor $\sim 2$ less light than our calculations and the one by FRV08. This difference is maximized around $z\sim 1$ up to a factor $\sim 4$.

Fig.~\ref{fig12} showed a comparison between our SFR density estimation and that predicted by SGPD10 as calculated by using the Eq.~\ref{star}. From $z\sim 3-1$ our SFR densities have different a behaviour: for our observational model increases up to $z=1$ and for SGPD10 keeps constant down to $z\sim 1.7$. For lower redshifts both models decline down to the local universe.

In general a very good agreement between the local EBL from the UV up to the mid-IR predicted by our method and the SAM of SGPD10-GSPD10 was seen in Fig.~\ref{fig13}. A factor $\sim 1.5$ higher intensity is found in the local UV from SAMs, and around the same factor lower intensity around 15~$\mu$m. For the far-IR peak, the difference comes from the different templates used for the dust component in the far-IR, which is given by the GRASIL code in the case of the SWIRE templates (which we use), and by a interpolation between the observed 70 and 160~$\mu$m by MIPS in the case of the templates used by these SAMs (\citealt{rieke09}).

The agreement on the evolution is very good as well as seen in Fig.~\ref{fig14}, even though at high redshift ($z>2$), GSPD10 predicts a factor of several more light in the UV. This is due to galaxies within the faint end of the theoretical LF at $z>2$. We recall that our observational model seems to already be overproducing light in the $B$-band for $z>1.5$ according to data in \citet{dahlen05} (see Fig.~\ref{fig11}).

\subsection{Overview on the cosmological picture}
\label{sec6.3}
The evolutionary path that we have in mind to interpret the evolution of the galaxy SED-type fractions is the following: AGNs are mostly formed by mergers between galaxies during hierarchical growth of dark matter halos (\eg \citealt{hopkins09}). When the merging galaxies are gas rich (also known as wet mergers), inflows of gas are produced leading to starburst galaxies and to the mass growth of the central black hole. The central black hole activity begins to expel the gas. Eventually, the gas is exhausted, switching off the AGN. The galaxy continues forming stars as a star-forming galaxy until the gas is fully depleted, then becomes a quiescent galaxy (\citealt{hopkins08a}; \citealt{hopkins08b}). 

It is now a well known observational fact that galaxies are bimodal in some properties such as colours (\citealt{strateva01}; \citealt{blanton03b}). They group in two different regions in colour-magnitude diagrams defining the red sequence and the blue cloud. Galaxies forming stars are in the blue cloud. Some galaxies have their star formation quenched when they become satellite galaxies in a larger halo, they cease to accrete gas, and they join the red sequence. Central galaxies form in the blue cloud, but they join the red sequence when they form a supermassive black hole and/or their halo mass exceeds approximately 10$^{12}$M$_{\sun}$ and/or they become satellite galaxies in a cluster. The most massive red galaxies cannot have simply be quenched central blue galaxies, since the latter are not massive enough; thus they must have been created by mergers without much star formation (also known as dry mergers). This effect is shown in Fig.~\ref{fig4}, taking into account that the $K$-band absolute magnitude $M_{K}$ is a good tracer of the galaxy stellar mass, as shown in \citet{brinchmann00}.

Massive galaxies today (very bright $M_{K}$) form their stars first, which is known as \emph{downsizing} (\citealt{cowie96}; \citealt{perezgonzalez08}). This initially seemed at odds with the hierarchical nature of the $\Lambda$CDM paradigm, in which small halos form first and agglomerate into larger ones. But the idea that star formation is efficient only in dark matter halos with a narrow range of masses naturally explains how the phenomenon of downsizing arises: halos that are massive today passed through the star-forming mass band between 10$^{8}$-10$^{12}$M$_{\odot}$ earlier and thus formed their stars earlier than halos that are less massive today (\citealt{croton09}; \citealt{conroy09}).

A careful examination of Fig.~\ref{fig4} reveals some interesting trends. We find that all the oldest galaxies ($\sim 13$~Gyr old, Ell13) are in the red sequence. However, the younger quiescent galaxies ($\sim 5$~Gyr old, Ell5) can be found in the red sequence as well as in the green valley (the region between the red sequence and the blue cloud). For the youngest quiescent galaxies ($\sim 2$~Gyr old, Ell2) we find that for $z>0.6$ they populate the green valley, while for $z<0.6$ they belong to the blue cloud. All the early-type star-forming galaxies (S0, Sa) are in the red sequence. Later-type star-forming galaxies such as Sb and Sc start to populate the green valley as well as the red sequence. Most of the very late-type star-forming galaxies (Sd, Sdm, Spi4) populate the blue cloud. Starburst galaxies are mainly in the green valley, but some of them are in the bluer region of the red sequence and in the redder region of the blue cloud. The same happens to AGNs, but they tend to be in the blue cloud more than in the red sequence.

We note that the increasing rate of quiescent galaxies as $z$ declines is roughly the same as the decreasing rate of starburst-type galaxies from $z\sim0.9-0.7$. One possible explanation would be the direct transformation of starbursts (either merger or huge-cold-gas reservoir triggered) directly to quiescent galaxies, without an intervening stage of significant star formation. Another explanation is that the characteristic time in which starburst-like galaxies consume their cold gas is the same that in which star-forming galaxies consume their lower cold-gas reservoir.  Thus the specific SFRs of these populations are different, but the rate at which starbursts enter the star-forming sequence is the same as the rate at which star-forming galaxies become quiescent. From $z\sim0.7-0.3$ the fraction of starbursts is very low, so the constant increase of the red sequence is modeled as due to AGNs preventing gas from cooling and forming stars.

\section{Summary and conclusions}
\label{summary}
A novel, robust and powerful method based on observations to derive the evolving spectrum of the extragalactic background light (EBL) between 0.1-1000~$\mu$m was presented. This model is based on the observed rest-frame $K$-band galaxy luminosity function (LF) over redshift by \citet{cirasuolo10} (C10), combined with an estimation of galaxy SED-type fractions based on a multiwavelength sample of $\sim 6000$ galaxies from AEGIS. This model has the following main advantages over other existing EBL models: transparent methodology, reproducibility, and utilizing direct galaxy data. The best available data sets are used (C10's LF and the AEGIS galaxy catalogue) observed over a wide redshift range. The galaxy evolution is directly observed in the rest-frame $K$-band up to $z=4$. Observed galaxies up to $z=1$ from the UV up to 24~$\mu$m with spectral energy distributions (SEDs) of 25 different types (from quiescent to rapidly star-forming galaxies, and including AGN galaxies) are taken into account in the same observational framework. A study of the uncertainties to the model directly from the data (such as uncertainties in the Schechter parameters of the C10 LF and the errors in the photometric catalogue) was done, and their propagated uncertainties to the $\gamma$-ray attenuation were studied.

A brief comparison with results from other recent EBL models is made here: Stecker et al. (2006) estimate a local EBL in the UV a factor of several higher than us, and in contradiction with recent $\gamma$-ray observations (\citealt{abdo10b}). A comparison with the FRV08's results was throughly presented through Sec.~\ref{results}. In general, our results are in good agreement, despite the fact that our modelings are different. Finke et al. (2010) have five different models based on different parametrizations of the SFR density of the universe and IMFs. The local EBL from the UV to the near-IR are similar to ours for their models \emph{C} and \emph{E}. \citet{kneiske10} claim to model a strict lower limit for the EBL. However our results for the local EBL in the UV are lower than the calculation by \citet{kneiske10}, but are in agreement with the robust lower limits from galaxy counts in the UV by \citet{madau00} and \citet{xu05}. In the near-IR, the model by \citet{kneiske10} is not compatible with the lower limits by \citet{keenan10}. Our observationally-based approach was also throughly compared with the results from the semi-analytic model (SAM) of galaxy formation by SGPD10 and GSPD10 in Sec.~\ref{SAMs}. Our EBL results are in general in good agreement at least for $z<2$, even though this SAM predicts more light (by up to a factor of several) than our observational approach in the UV, and a factor $\sim2-3$ less light in the far-IR.

Our methodology provides a tool for calculating the EBL more accurately at the longest wavelengths when a better understanding on the far-IR galaxy SEDs, new photometry, and deeper LFs at those wavelengths are available from the Herschel Space Observatory. 

Two extrapolations of the galaxy SED-type fractions to $z>1$ were considered, showing that these assumptions only affect the far-IR. It was calculated that the population with SED features of quiescent local galaxies increases a factor $\sim 2$ since $z\sim1$. The star-forming population remains roughly constant, while the starburst-like population decreases very quickly from around $\sim20\%$ at $z\sim1$. The AGN-like population decreases slower than the starburst-like population from almost 20\% at $z\sim1$ to just around 2\% at $z\sim0.3$. Data from the future James Webb Space Telescope will help to determine the galaxy SED-type fractions at $z>1$.

A low intensity local EBL ($z=0$) was found, matching the lower limits from galaxy counts up to $\sim30$~$\mu$m. For longer wavelengths, our model predicts higher intensities than the data from galaxy counts, in agreement with direct measurements.

Our results are also compatible with all the upper limits from $\gamma$-ray astronomy according to the standard framework for the propagation of VHE photons through the universe, even though to account for the highest energies detected by \citet{aharonian99} for Mrk~501 we have to assume a $\Gamma_{int}<1.5$, appeal to statistical and systematic uncertainties on this VHE spectrum, or attenuation uncertainties due to uncertainties in the EBL for such high energies as discussed in Sec.\ref{attenuation}.

The EBL uncertainties in far-IR leading to attenuation uncertainties of a factor of several for energies larger than $\sim 10$~TeV needs to be addressed by the current and next generation of IR telescopes providing new photometric data and a better understanding of the galaxy IR emission. $\gamma$-ray astronomy may constrain these uncertainties from a better understanding of the emission mechanisms at those high energies (helped with simultaneous multiwavelength observations) and of the instrumental systematic uncertainties. Observations aimed to measure photons with energies higher than $\sim 10$~TeV at $z<0.3$ are encouraged.

It is worth mentioning that high energy (30~MeV-30~GeV) $\gamma$-rays are detected by Fermi for $z\le2.5$ from AGNs (\citealt{abdo10}) and for $z\le4.5$ (\citealt{abdo09}) from GRBs. The reasons for these high redshift detections include a larger $\gamma$-ray flux at lower energies and a lower density of EBL target photons that can interact with these $\gamma$-rays.  Understanding the evolution of the EBL at UV wavelengths is essential to interpreting observations of these high-redshift sources. New observations of AGNs as well as the first GRB detection in the VHE range would help to make new and stronger constraints on the EBL; see \citet{gilmore10a}.

The universe, according to our observationally-based model, is more transparent than the estimation from FRV08 (a factor $\sim2$ in flux) for VHE photons coming from low-redshift sources ($z\sim0.1$) for energies between $\sim6-15$~TeV, but still the uncertainties here from the EBL modeling are large (a factor of several). The same attenuation than FRV08 is estimated for other energies. For VHE photons coming from larger-redshift sources ($z\sim1$), roughly the same attenuation as FRV08 is estimated. Here the attenuation uncertainties (for energies available to $\gamma$-ray telescopes) due to the uncertainties on the EBL modeling are low in comparison with other effects. At these redshifts the uncertainties on the EBL-corrected spectra are dominated by instrumental systematic uncertainties. We may conclude that it is not expected to observe any such high redshift ($z\sim1$) multi-TeV $\gamma$-ray photons from blazars with the current or even next telescope generation such as the Cherenkov Telescope Array (CTA, \citealt{doro09}) or the Advanced Gamma-ray Imaging System (AGIS, \citealt{buckley08}), but we indeed expect a promising future for sub-TeV detections at these high redshifts.


\section*{Acknowledgments}
We are grateful to Ranga-Ram Chary, Darren Croton, Amy Furniss, Valentino Gonz\'alez, Patrik Jonsson, Nepomuk Otte, Miguel \'Angel S\'anchez-Conde, David Williams and Li Yan-Rong for fruitful discussions, and to the anonymous referee for useful comments. We also thank Valentino Gonz\'alez for his help running FAST. A.D. warmly thanks for their hospitality the staff at SCIPP and the Astronomy Department-UCSC, where most of this work was done. A.D.'s work has been supported by the Spanish Ministerio de Educaci\'on y Ciencia and the European regional development fund (FEDER) under projects FIS2008-04189 and CPAN-Ingenio (CSD2007-00042), and by the Junta de Andaluc\'{\i}a (P07-FQM-02894). A.D. also acknowledges the Spanish MEC for a FPI grant. J.P.'s and R.G.'s research is supported by Fermi Theory Grants NNX08AW37G, NNX09AT98G and NNX10AP54G. D.R. acknowledges the support of the National Science Foundation through grants AST-0507483 and AST-0808133. F.P. thanks the support of the Spanish MICINN's Consolider-Ingenio 2010 Programme under grant MultiDark CSD2009-00064. R.G. was also supported by a research fellowship from the SISSA Astrophysics Sector. This study makes use of data from AEGIS, a multiwavelength sky survey conducted with the Chandra, GALEX, Hubble, Keck, CFHT, MMT, Subaru, Palomar, Spitzer, VLA, and other telescopes and supported in part by the NSF, NASA, and the STFC.

\bibliographystyle{plain}

\begin{thebibliography}{}

\bibitem[\protect\citeauthoryear{Abdo et al.}{2009}]{abdo09} Abdo A.~A., et al., 2009, Sci, 323, 1688

\bibitem[\protect\citeauthoryear{Abdo et al.}{2010a}]{abdo10} Abdo A.~A., et al., 2010a, ApJ, 715, 429

\bibitem[\protect\citeauthoryear{Abdo et al.}{2010b}]{abdo10b} Abdo A.~A., et al., 2010b, arXiv, arXiv:1005.0996

\bibitem[\protect\citeauthoryear{Acciari et 
al.}{2009}]{veritas09} Acciari V.~A., et al., 2009, ApJ, 693, 
L104

\bibitem[\protect\citeauthoryear{Aharonian et 
al.}{1999}]{aharonian99} Aharonian F., et al., 1999, A\&A, 349, 11 

\bibitem[\protect\citeauthoryear{Aharonian et 
al.}{2001}]{aharonian01} Aharonian F.~A., et al., 2001, A\&A, 366, 62 

\bibitem[\protect\citeauthoryear{Aharonian et 
al.}{2006}]{aharonian06} Aharonian F., et al., 2006, Nature, 440, 
1018

\bibitem[\protect\citeauthoryear{Aharonian et 
al.}{2007}]{aharonian07c} Aharonian F., et al., 2007, A\&A, 475, L9

\bibitem[\protect\citeauthoryear{Aharonian et 
al.}{2008}]{aharonian08} Aharonian, F.~A., Khangulyan, D., \& Costamante, L \ 2008, MNRAS, 387, 1206

\bibitem[\protect\citeauthoryear{Albert et al.}{2007}]{albert07} Albert J., et al., 2007, ApJ, 654, L119 

\bibitem[\protect\citeauthoryear{Albert et al.}{2008}]{albert08} Albert J., et al., 2008, Sci, 320, 1752

\bibitem[\protect\citeauthoryear{Arnouts et 
al.}{2002}]{arnouts02} Arnouts S., et al., 2002, MNRAS, 329, 355

\bibitem[\protect\citeauthoryear{Arnouts et 
al.}{2007}]{arnouts07} Arnouts S., et al., 2007, A\&A, 476, 137 

\bibitem[\protect\citeauthoryear{Barmby et al.}{2008}]{barmby08} 
Barmby P., Huang J.-S., Ashby M.~L.~N., Eisenhardt P.~R.~M., Fazio G.~G., 
Willner S.~P., Wright E.~L., 2008, ApJS, 177, 431

\bibitem[\protect\citeauthoryear{Barro et 
al.}{2009}]{barro09} Barro G., et al., 2009, A\&A, 494, 63

\bibitem[\protect\citeauthoryear{Bell et al.}{2003}]{bell03} 
Bell E.~F., McIntosh D.~H., Katz N., Weinberg M.~D., 2003, ApJS, 149, 289

\bibitem[\protect\citeauthoryear{Bernstein}{2007}]{bernstein07} 
Bernstein R.~A., 2007, ApJ, 666, 663

\bibitem[\protect\citeauthoryear{Berta et 
al.}{2010}]{berta10} Berta S., et al., 2010, A\&A, 518, L30 

\bibitem[\protect\citeauthoryear{B{\'e}thermin et al.}{2010}]{bethermin10} B{\'e}thermin M., Dole H., Beelen A., Aussel H., 2010, A\&A, 512, A78

\bibitem[\protect\citeauthoryear{Blanton et 
al.}{2003a}]{blanton03a} Blanton M.~R., et al., 2003a, ApJ, 592, 819 

\bibitem[\protect\citeauthoryear{Blanton et 
al.}{2003b}]{blanton03b} Blanton M.~R., et al., 2003b, ApJ, 594, 186 

\bibitem[\protect\citeauthoryear{Blanton}{2006}]{blanton06} 
Blanton M.~R., 2006, ApJ, 648, 268

\bibitem[\protect\citeauthoryear{B{\"o}ttcher}{2007}]{boettcher07} B{\"o}ttcher M., 2007, Ap\&SS, 309, 95

\bibitem[\protect\citeauthoryear{B{\"o}ttcher, Dermer 
\& Finke}{2008}]{bottcher08} B{\"o}ttcher M., Dermer C.~D., Finke J.~D., 2008, ApJ, 679, L9

\bibitem[\protect\citeauthoryear{Bramel et al.}{2005}]{bramel05} 
Bramel D.~A., et al., 2005, ApJ, 629, 108 

\bibitem[\protect\citeauthoryear{Brinchmann 
\& Ellis}{2000}]{brinchmann00} Brinchmann J., Ellis R.~S., 2000, ApJ, 536, L77 

\bibitem[\protect\citeauthoryear{Bruzual 
\& Charlot}{2003}]{bruzual03} Bruzual G., Charlot S., 2003, MNRAS, 344, 1000 

\bibitem[\protect\citeauthoryear{Buckley et 
al.}{2008}]{buckley08} Buckley J., et al., 2008, AIPC, 1085, 902 

\bibitem[\protect\citeauthoryear{Calzetti et 
al.}{2000}]{calzetti00} Calzetti D., Armus L., Bohlin R.~C., 
Kinney A.~L., Koornneef J., Storchi-Bergmann T., 2000, ApJ, 533, 682

\bibitem[\protect\citeauthoryear{Cambr{\'e}sy et al.}{2001}]{cambresy01} Cambr{\'e}sy L., Reach W.~T., Beichman  C.~A., Jarrett T.~H., 2001, ApJ, 555, 563 

\bibitem[\protect\citeauthoryear{Cameron et 
al.}{2009}]{cameron09} Cameron E., Driver S.~P., Graham A.~W., 
Liske J., 2009, ApJ, 699, 105 

\bibitem[\protect\citeauthoryear{Chabrier}{2003}]{chabrier03} 
Chabrier G., 2003, PASP, 115, 763 

\bibitem[\protect\citeauthoryear{Chary et al.}{2004}]{chary04} 
Chary R., et al., 2004, ApJS, 154, 80 

\bibitem[\protect\citeauthoryear{Chary 
\& Pope}{2010}]{chary10} Chary R.-R., Pope A., 2010, arXiv, arXiv:1003.1731 

\bibitem[\protect\citeauthoryear{Cirasuolo et 
al.}{2010}]{cirasuolo10} Cirasuolo M., McLure R.~J., Dunlop J.~S., 
Almaini O., Foucaud S., Simpson C., 2010, MNRAS, 401, 1166 (C10)

\bibitem[\protect\citeauthoryear{Coil et al.}{2004}]{coil04} 
Coil A.~L., Newman J.~A., Kaiser N., Davis M., Ma C.-P., Kocevski D.~D., 
Koo D.~C., 2004, ApJ, 617, 765

\bibitem[\protect\citeauthoryear{Conroy 
\& Wechsler}{2009}]{conroy09} Conroy C., Wechsler R.~H., 2009, ApJ, 696, 620

\bibitem[\protect\citeauthoryear{Conselice et al.}{2008}]{conselice08} Conselice C.~J., Bundy K., U V., Eisenhardt P., Lotz J., Newman J., 2008, MNRAS, 383, 1366

\bibitem[\protect\citeauthoryear{Cowie et al.}{1996}]{cowie96} 
Cowie L.~L., Songaila A., Hu E.~M., Cohen J.~G., 1996, AJ, 112, 839 

\bibitem[\protect\citeauthoryear{Croton}{2009}]{croton09} Croton 
D.~J., 2009, MNRAS, 394, 1109

\bibitem[\protect\citeauthoryear{Cuillandre et 
al.}{2001}]{cuillandre01} Cuillandre J.-C., Luppino G., Starr B., 
Isani S., 2001, sf2a.conf, 605 

\bibitem[\protect\citeauthoryear{Daddi et al.}{2007}]{daddi07} 
Daddi E., et al., 2007, ApJ, 670, 156 

\bibitem[\protect\citeauthoryear{Dahlen et al.}{2005}]{dahlen05} 
Dahlen T., Mobasher B., Somerville R.~S., Moustakas L.~A., Dickinson M., 
Ferguson H.~C., Giavalisco M., 2005, ApJ, 631, 126 

\bibitem[\protect\citeauthoryear{Dahlen et al.}{2007}]{dahlen07} 
Dahlen T., Mobasher B., Dickinson M., Ferguson H.~C., Giavalisco M., 
Kretchmer C., Ravindranath S., 2007, ApJ, 654, 172 

\bibitem[\protect\citeauthoryear{Danforth et 
al.}{2010}]{danforth10} Danforth C.~W., Keeney B.~A., Stocke 
J.~T., Shull J.~M., Yao Y., 2010, ApJ, 720, 976

\bibitem[\protect\citeauthoryear{Davis et al.}{2007}]{davis07} 
Davis M., et al., 2007, ApJ, 660, L1

\bibitem[\protect\citeauthoryear{Devlin et al.}{2009}]{devlin09} 
Devlin M.~J., et al., 2009, Natur, 458, 737 

\bibitem[\protect\citeauthoryear{Dickinson 
et al.}{2007}]{dickinson07} Dickinson M., FIDEL team, 2007, AAS, 38, 822 

\bibitem[\protect\citeauthoryear{Doro}{2009}]{doro09} Doro M. for the CTA consortium, 2009, arXiv, arXiv:0908.1410

\bibitem[\protect\citeauthoryear{Driver et al.}{2008}]{driver08} Driver S.~P., Popescu C.~C., Tuffs R.~J., Graham A.~W., Liske J., Baldry I., 2008, ApJ, 678, L101

\bibitem[\protect\citeauthoryear{Dwek 
\& Krennrich}{2005}]{dwek05} Dwek E., Krennrich F., 2005, ApJ, 618, 657

\bibitem[\protect\citeauthoryear{Faber et al.}{2003}]{faber03} 
Faber S.~M., et al., 2003, SPIE, 4841, 1657

\bibitem[\protect\citeauthoryear{Faber et al.}{2007}]{faber07} 
Faber S.~M., et al., 2007, ApJ, 665, 265 

\bibitem[\protect\citeauthoryear{Fardal et al.}{2007}]{fardal07} Fardal M.~A., Katz N., Weinberg D.~H., Dav{\'e} R., 2007, MNRAS, 379, 985

\bibitem[\protect\citeauthoryear{Fazio et al.}{2004}]{fazio04} 
Fazio G.~G., et al., 2004, ApJS, 154, 39 

\bibitem[\protect\citeauthoryear{Finkbeiner, Davis 
\& Schlegel}{2000}]{finkbeiner00} Finkbeiner D.~P., Davis M., Schlegel D.~J., 2000, ApJ, 544, 81 

\bibitem[\protect\citeauthoryear{Finke, Razzaque \& Dermer}{2010}]{finke10} Finke J.~D., Razzaque S., Dermer C.~D., 2010, ApJ, 712, 238

\bibitem[\protect\citeauthoryear{Franceschini, Rodighiero 
\& Vaccari}{2008}]{franceschini08} Franceschini A., Rodighiero G., Vaccari M., 2008, A\&A, 487, 837 (FRV08)

\bibitem[\protect\citeauthoryear{Frayer et al.}{2006}]{frayer06} 
Frayer D.~T., et al., 2006, ApJ, 647, L9

\bibitem[\protect\citeauthoryear{Gabasch et 
al.}{2004}]{gabasch04} Gabasch A., et al., 2004, A\&A, 421, 41

\bibitem[\protect\citeauthoryear{Gabasch et 
al.}{2006}]{gabasch06} Gabasch A., et al., 2006, A\&A, 448, 101

\bibitem[\protect\citeauthoryear{Gardner, Brown \& Ferguson}{2000}]{gardner00} Gardner J.~P., Brown T.~M., Ferguson H.~C., 2000, ApJ, 542, L79 

\bibitem[\protect\citeauthoryear{Gehrels 
\& Michelson}{1999}]{gehrels99} Gehrels N., Michelson P., 1999, APh, 11, 277

\bibitem[\protect\citeauthoryear{Gilmore, Prada \& Primack}{2010}]{gilmore10a} Gilmore R.~C., Prada F., Primack J., 2010, MNRAS, 402, 565 

\bibitem[\protect\citeauthoryear{Gilmore et al.}{2010}]{gilmore10b} Gilmore R.~C., Somerville R.~S., Primack J.~R., Dom\'inguez A., 2010, in prep. (GSPD10)

\bibitem[\protect\citeauthoryear{Gorjian, Wright 
\& Chary}{2000}]{gorjian00} Gorjian V., Wright E.~L., Chary R.~R., 2000, ApJ, 536, 550 

\bibitem[\protect\citeauthoryear{Goto et al.}{2003}]{goto03} 
Goto T., Yamauchi C., Fujita Y., Okamura S., Sekiguchi M., Smail I., 
Bernardi M., Gomez P.~L., 2003, MNRAS, 346, 601

\bibitem[\protect\citeauthoryear{Gould 
\& Schr{\'e}der}{1966}]{gould66} Gould R.~J., Schr{\'e}der G., 1966, PhRvL, 16, 252

\bibitem[\protect\citeauthoryear{Hartman et 
al.}{1999}]{hartman99} Hartman R.~C., et al., 1999, ApJS, 123, 79 

\bibitem[\protect\citeauthoryear{Hauser et al.}{1998}]{hauser98} 
Hauser M.~G., et al., 1998, ApJ, 508, 25 

\bibitem[\protect\citeauthoryear{Hauser 
\& Dwek}{2001}]{hauser01} Hauser M.~G., Dwek E., 2001, ARA\&A, 39, 249

\bibitem[\protect\citeauthoryear{Hinton}{2004}]{hinton04} Hinton 
J.~A., 2004, NewAR, 48, 331

\bibitem[\protect\citeauthoryear{Hopkins et 
al.}{2008a}]{hopkins08a} Hopkins P.~F., Hernquist L., Cox T.~J., 
Kere{\v s} D., 2008a, ApJS, 175, 356 

\bibitem[\protect\citeauthoryear{Hopkins et 
al.}{2008b}]{hopkins08b} Hopkins P.~F., Cox T.~J., Kere{\v s} D., 
Hernquist L., 2008b, ApJS, 175, 390 

\bibitem[\protect\citeauthoryear{Hopkins et 
al.}{2009}]{hopkins09} Hopkins P.~F., et al., 2009, MNRAS, 397, 
802 

\bibitem[\protect\citeauthoryear{Hopwood et 
al.}{2010}]{hopwood10} Hopwood R., et al., 2010, ApJ, 716, L45

\bibitem[\protect\citeauthoryear{Huang et al.}{2007}]{huang07}
Huang J.-S., et al., 2007, ApJ, 664, 840 

\bibitem[\protect\citeauthoryear{Huang et al.}{2010}]{huang10} Huang J.-S., et al., 2010, in prep.

\bibitem[\protect\citeauthoryear{Ilbert et 
al.}{2005}]{ilbert05} Ilbert O., et al., 2005, A\&A, 439, 863 

\bibitem[\protect\citeauthoryear{Ilbert et 
al.}{2006}]{ilbert06} Ilbert O., et al., 2006, A\&A, 457, 841

\bibitem[\protect\citeauthoryear{Jones et al.}{2006}]{jones06} 
Jones D.~H., Peterson B.~A., Colless M., Saunders W., 2006, MNRAS, 369, 25 

\bibitem[\protect\citeauthoryear{Katarzy{\'n}ski et 
al.}{2006}]{katarzynski06} Katarzy{\'n}ski K., Ghisellini G., 
Tavecchio F., Gracia J., Maraschi L., 2006, MNRAS, 368, L52

\bibitem[\protect\citeauthoryear{Keenan et al.}{2010}]{keenan10} Keenan R.~C., Barger A.~J., Cowie L.~L., Wang W.-H., 2010, arXiv, arXiv:1008.4216 

\bibitem[\protect\citeauthoryear{Kennicutt}{1998}]{kennicutt98} Kennicutt R.~C., Jr., 1998, ARA\&A, 36, 189 

\bibitem[\protect\citeauthoryear{Kneiske, Mannheim 
\& Hartmann}{2002}]{kneiske02} Kneiske T.~M., Mannheim K., Hartmann D.~H., 2002, A\&A, 386, 1

\bibitem[\protect\citeauthoryear{Kneiske \& Dole}{2010}]{kneiske10} Kneiske T.~M., Dole H., 2010, A\&A, 515, A19

\bibitem[\protect\citeauthoryear{Kochanek et 
al.}{2001}]{kochanek01} Kochanek C.~S., et al., 2001, ApJ, 560, 
566 

\bibitem[\protect\citeauthoryear{Kriek et al.}{2009}]{kriek09} 
Kriek M., van Dokkum P.~G., Labb{\'e} I., Franx M., Illingworth G.~D., 
Marchesini D., Quadri R.~F., 2009, ApJ, 700, 221 

\bibitem[\protect\citeauthoryear{Lagache et al.}{2000}]{lagache00b} Lagache G., Haffner L.~M., Reynolds R.~J., Tufte S.~L., 2000, A\&A, 354, 247

\bibitem[\protect\citeauthoryear{Lagache, Puget \& Dole}{2005}]{lagache05} Lagache G., Puget J.-L., Dole H., 2005, ARA\&A, 43, 727

\bibitem[\protect\citeauthoryear{Lawrence et 
al.}{2007}]{lawrence07} Lawrence A., et al., 2007, MNRAS, 379, 
1599 

\bibitem[\protect\citeauthoryear{LeFloc'h et 
al.}{2009}]{lefloch09} LeFloc'h E., et al., 2009, ApJ, 703, 222 

\bibitem[\protect\citeauthoryear{Levenson 
\& Wright}{2008}]{levenson08} Levenson L.~R., Wright E.~L., 2008, ApJ, 683, 585

\bibitem[\protect\citeauthoryear{Lorenz}{2004}]{lorenz04} Lorenz 
E., 2004, NewAR, 48, 339

\bibitem[\protect\citeauthoryear{Madau 
\& Phinney}{1996}]{madau96} Madau P., Phinney E.~S., 1996, ApJ, 456, 124 

\bibitem[\protect\citeauthoryear{Madau 
\& Pozzetti}{2000}]{madau00} Madau P., Pozzetti L., 2000, MNRAS, 312, L9

\bibitem[\protect\citeauthoryear{Magnelli et 
al.}{2009}]{magnelli09} Magnelli B., Elbaz D., Chary R.~R., Dickinson M., Le Borgne D., Frayer D.~T., Willmer C.~N.~A., 2009, A\&A, 496, 57

\bibitem[\protect\citeauthoryear{Malkan 
\& Stecker}{1998}]{malkan98} Malkan M.~A., Stecker F.~W., 1998, ApJ, 496, 13 

\bibitem[\protect\citeauthoryear{Marchesini et 
al.}{2007}]{marchesini07} Marchesini D., et al., 2007, ApJ, 656, 42 

\bibitem[\protect\citeauthoryear{Matsumoto et 
al.}{2005}]{matsumoto05} Matsumoto T., et al., 2005, ApJ, 626, 31

\bibitem[\protect\citeauthoryear{Matsuura et 
al.}{2010}]{matsuura10} Matsuura S., et al., 2010, arXiv, arXiv:1002.3674 

\bibitem[\protect\citeauthoryear{Mattila}{2006}]{mattila06} Mattila K., 2006, MNRAS, 372, 1253 

\bibitem[\protect\citeauthoryear{Matute et al.}{2006}]{matute06} Matute I., La Franca F., Pozzi F., Gruppioni C., Lari C., Zamorani G., 2006, A\&A, 451, 443

\bibitem[\protect\citeauthoryear{Mazin 
\& Raue}{2007}]{mazin07} Mazin D., Raue M., 2007, A\&A, 471, 439

\bibitem[\protect\citeauthoryear{Men{\'e}ndez-Delmestre et 
al.}{2009}]{menendez-delmestre09} Men{\'e}ndez-Delmestre K., et al., 2009, 
ApJ, 699, 667

\bibitem[\protect\citeauthoryear{Metcalfe et 
al.}{2003}]{metcalfe03} Metcalfe L., et al., 2003, A\&A, 407, 791

\bibitem[\protect\citeauthoryear{Montero-Dorta 
\& Prada}{2009}]{montero-dorta09} Montero-Dorta A.~D., Prada F., 2009, MNRAS, 399, 1106 

\bibitem[\protect\citeauthoryear{Morrissey et 
al.}{2007}]{morrissey07} Morrissey P., et al., 2007, ApJS, 173, 682

\bibitem[\protect\citeauthoryear{Murphy et al.}{2009}]{murphy09} 
Murphy E.~J., Chary R.-R., Alexander D.~M., Dickinson M., Magnelli B., 
Morrison G., Pope A., Teplitz H.~I., 2009, ApJ, 698, 1380

\bibitem[\protect\citeauthoryear{Newman et al.}{2010}]{newman10} Newman J.~A., et al., 2010, in prep.

\bibitem[\protect\citeauthoryear{Nikishov}{1962}]{nikishov62} Nikishov A.~I., 1962, Sov. Phys. JETP, 14, 393

\bibitem[\protect\citeauthoryear{Norberg et 
al.}{2002}]{norberg02} Norberg P., et al., 2002, MNRAS, 336, 907 

\bibitem[\protect\citeauthoryear{Papovich et 
al.}{2004}]{papovich04} Papovich C., et al., 2004, ApJS, 154, 70

\bibitem[\protect\citeauthoryear{Peebles}{1993}]{peebles93} 
Peebles P.~J.~E., 1993, Principles of Physical Cosmology, Princeton University Press

\bibitem[\protect\citeauthoryear{P{\'e}rez-Gonz{\'a}lez et 
al.}{2008}]{perezgonzalez08} P{\'e}rez-Gonz{\'a}lez P.~G., et al., 
2008, ApJ, 675, 234

\bibitem[\protect\citeauthoryear{Polletta et 
al.}{2007}]{polletta07} Polletta M., et al., 2007, ApJ, 663, 81

\bibitem[\protect\citeauthoryear{Prandini et al.}{2010}]{prandini10} Prandini E., Bonnoli G., Maraschi L., Mariotti M., Tavecchio F., 2010, MNRAS, 405, L76 

\bibitem[\protect\citeauthoryear{Primack et 
al.}{1999}]{primack99} Primack J.~R., Bullock J.~S., Somerville 
R.~S., MacMinn D., 1999, APh, 11, 93

\bibitem[\protect\citeauthoryear{Primack et 
al.}{2001}]{primack01} Primack J.~R., Somerville R.~S., Bullock 
J.~S., Devriendt J.~E.~G., 2001, AIPC, 558, 463

\bibitem[\protect\citeauthoryear{Reddy et al.}{2005}]{reddy05} 
Reddy N.~A., Erb D.~K., Steidel C.~C., Shapley A.~E., Adelberger K.~L., 
Pettini M., 2005, ApJ, 633, 748 

\bibitem[\protect\citeauthoryear{Reddy et al.}{2006}]{reddy06} 
Reddy N.~A., Steidel C.~C., Fadda D., Yan L., Pettini M., Shapley A.~E., 
Erb D.~K., Adelberger K.~L., 2006, ApJ, 644, 792 

\bibitem[\protect\citeauthoryear{Rieke et al.}{2009}]{rieke09} 
Rieke G.~H., Alonso-Herrero A., Weiner B.~J., P{\'e}rez-Gonz{\'a}lez P.~G., 
Blaylock M., Donley J.~L., Marcillac D., 2009, ApJ, 692, 556 

\bibitem[\protect\citeauthoryear{Rodighiero et 
al.}{2010}]{rodighiero10} Rodighiero G., et al., 2010, A\&A, 515, A8

\bibitem[\protect\citeauthoryear{Salim et al.}{2009}]{salim09} 
Salim S., et al., 2009, ApJ, 700, 161 

\bibitem[\protect\citeauthoryear{Salpeter}{1955}]{salpeter55} Salpeter E.~E., 1955, ApJ, 121, 161 

\bibitem[\protect\citeauthoryear{S{\'a}nchez-Conde et 
al.}{2009}]{sanchezconde09} S{\'a}nchez-Conde M.~A., Paneque D., Bloom 
E., Prada F., Dom{\'{\i}}nguez A., 2009, PhRvD, 79, 123511 

\bibitem[\protect\citeauthoryear{Schawinski et 
al.}{2009}]{schawinski09} Schawinski K., et al., 2009, MNRAS, 396, 
818 

\bibitem[\protect\citeauthoryear{Schechter}{1976}]{schechter76} 
Schechter P., 1976, ApJ, 203, 297

\bibitem[\protect\citeauthoryear{Schlegel, Finkbeiner 
\& Davis}{1998}]{schlegel98} Schlegel D.~J., Finkbeiner D.~P., Davis M., 1998, ApJ, 500, 525 

\bibitem[\protect\citeauthoryear{Serjeant 
\& Harrison}{2005}]{serjeant05} Serjeant S., Harrison D., 2005, MNRAS, 356, 192

\bibitem[\protect\citeauthoryear{Sikora et al.}{2009}]{sikora09} 
Sikora M., Stawarz {\L}., Moderski R., Nalewajko K., Madejski G.~M., 2009, 
ApJ, 704, 38 

\bibitem[\protect\citeauthoryear{Silva, Maiolino \& Granato}{2004}]{silva04} Silva L., Maiolino R., Granato G.~L., 2004, MNRAS, 355, 973

\bibitem[\protect\citeauthoryear{Silva et al.}{1998}]{silva07} 
Silva L., Granato G.~L., Bressan A., Danese L., 1998, ApJ, 509, 103 

\bibitem[\protect\citeauthoryear{Skibba et al.}{2009}]{skibba09} 
Skibba R.~A., et al., 2009, MNRAS, 1243

\bibitem[\protect\citeauthoryear{Soifer 
\& Neugebauer}{1991}]{soifer91} Soifer B.~T., Neugebauer G., 1991, AJ, 101, 354

\bibitem[\protect\citeauthoryear{Somerville 
\& Primack}{1999}]{somerville99} Somerville R.~S., Primack J.~R., 1999, MNRAS, 310, 1087 

\bibitem[\protect\citeauthoryear{Somerville, Primack, 
\& Faber}{2001}]{somerville01} Somerville R.~S., Primack J.~R., Faber S.~M., 2001, MNRAS, 320, 504 

\bibitem[\protect\citeauthoryear{Somerville et 
al.}{2008}]{somerville08} Somerville R.~S., Hopkins P.~F., Cox 
T.~J., Robertson B.~E., Hernquist L., 2008, MNRAS, 391, 481

\bibitem[\protect\citeauthoryear{Somerville et al.}{2010}]{somerville10} Somerville R.~S., Gilmore R.~C, Primack J.~R, Dom\'inguez A., 2010, in prep. (SGPD10)

\bibitem[\protect\citeauthoryear{Stecker 
\& Glashow}{2001}]{stecker01} Stecker F.~W., Glashow S.~L., 2001, APh, 16, 97

\bibitem[\protect\citeauthoryear{Stecker, Malkan \& Scully}{2006}]{stecker06} Stecker F.~W., Malkan M.~A., Scully S.~T., 2006, ApJ, 648, 774

\bibitem[\protect\citeauthoryear{Stecker, Baring 
\& Summerlin}{2007}]{stecker07} Stecker F.~W., Baring M.~G., Summerlin E.~J., 2007, ApJ, 667, L29

\bibitem[\protect\citeauthoryear{Strateva et 
al.}{2001}]{strateva01} Strateva I., et al., 2001, AJ, 122, 1861 

\bibitem[\protect\citeauthoryear{Takeuchi et 
al.}{2006}]{takeuchi06} Takeuchi T.~T., Ishii T.~T., Dole H., Dennefeld M., Lagache G., Puget J.-L., 2006, A\&A, 448, 525

\bibitem[\protect\citeauthoryear{Taylor et al.}{2009}]{taylor09} 
Taylor E.~N., et al., 2009, ApJ, 694, 1171

\bibitem[\protect\citeauthoryear{Yang 
\& Wang}{2010}]{yang10} Yang J., Wang J., 2010, PASJ, 62, L23 

\bibitem[\protect\citeauthoryear{Younger 
\& Hopkins}{2010}]{younger10} Younger J.~D., Hopkins P.~F., 2010, arXiv, arXiv:1003.4733 

\bibitem[\protect\citeauthoryear{Xu et al.}{2005}]{xu05} Xu 
C.~K., et al., 2005, ApJ, 619, L11

\bibitem[\protect\citeauthoryear{Weekes et al.}{2002}]{weekes02} 
Weekes T.~C., et al., 2002, APh, 17, 221

\bibitem[\protect\citeauthoryear{Willmer et 
al.}{2006}]{willmer06} Willmer C.~N.~A., et al., 2006, ApJ, 647, 
853 

\bibitem[\protect\citeauthoryear{Wilson et al.}{2003}]{wilson03} 
Wilson J.~C., et al., 2003, SPIE, 4841, 451 

\bibitem[\protect\citeauthoryear{Wyder et al.}{2005}]{wyder05} 
Wyder T.~K., et al., 2005, ApJ, 619, L15

\bibitem[\protect\citeauthoryear{Wuyts et al.}{2009}]{wuyts09} 
Wuyts S., et al., 2009, ApJ, 700, 799 

\end{thebibliography}

\label{lastpage}
\end{document}